\documentclass[prx,twocolumn,amsmath,amssymb]{revtex4}

\usepackage{color}
\usepackage{bm}
\usepackage{dcolumn}
\usepackage[dvips]{graphicx}
\usepackage{slashed}

\usepackage{ulem}

\newcommand{\1}{\mbox{1}\hspace{-0.25em}\mbox{l}}

\newcommand{\tr}{{\rm tr}\,}

\begin{document}

\title{
	Edge states,  corner states,  and flat bands in a two-dimensional $\cal PT$-symmetric system
}

\author{Akira Yoshida, Yuria Otaki, Rimako Otaki, and Takahiro Fukui}
\affiliation{Department of Physics, Ibaraki University, Mito 310-8512, Japan}

\date{\today}

\begin{abstract}	 
	We study corner states on a flat band in the square lattice. To this end, we introduce
	a two dimensional model including  Su-Schrieffer-Heeger  type bond alternation responsible for corner states
	as well as
	next-nearest neighbor hoppings  yielding flat bands.
	The key symmetry of the model for corner states is space-time inversion ($\cal PT$) symmetry, which guarantees quantized Berry phases.
	This implies that edge states as well as corner states would show up if boundaries are introduced to
	the system.
	We also argue that an infinitesimal $\cal PT$ symmetry-breaking perturbation could drive flat bands into flat Chern bands.
\end{abstract}

\pacs{
}

\maketitle

\section{Introduction}

Topological properties of gapped ground states for bulk systems
could be unveiled  by edge states if boundaries are introduced to those systems.
This is called bulk-edge (boundary) correspondence \cite{Hatsugai:1993fk}, which is nowadays
one of the fundamental concepts in condensed matter physics.
Except for the quantum Hall effect (QHE) states \cite{Thouless:1982uq,kohmoto:85},
bulk topological invariants in various classes of topological insulators
\cite{Kane:2005aa,Qi:2008aa,Schnyder:2008aa}
are not necessarily related with the observables such as the Hall conductance.
Therefore, experiments on topological insulators are mainly based on the observation of gapless surface states
\cite{Konig:1977fk,Hasan:2010fk,Qi:2011kx,Ando:2013aa}.

The conventional bulk-edge correspondence is the relationship between $d$-dimensional bulk gapped ground states
and their $d-1$ dimensional surface gapless states.
Recently, the bulk-edge correspondence has been extended to
higher-order topological insulators (HOTI)
\cite{Slager:2015aa,Benalcazar:2017aa,Benalcazar:2017ab,Liu:2017aa,Schindler:2018ab}
which have $d-D$ dimensional boundary states generically with $D>1$.
The HOTI have been studied extensively with a broad interest in, e.g.,
symmetry properties and classification
\cite{Langbehn:2017aa,Song:2017aa,Khalaf:2018cr,Fukui:2018aa,Trifunovic:2019aa,Benalcazar:2019aa,Matsugatani:2018aa},
model construction \cite{Ezawa:2018aa,Ezawa:2018ab,Calugaru:2019aa,Ezawa:2018ac},
a field theoretical point of view \cite{Hashimoto:2017aa},  superconducting systems
\cite{Wang:2018aa,Hsu:2018aa,Ghorashi:2019aa},
interaction effects \cite{You:2018aa}, Floquet systems \cite{Bomantara:2019aa,1811.04808}, etc.
Among them, the HOTI on the breathing Kagome lattice \cite{Ezawa:2018aa} may be quite interesting,
since the model shows a flat band as well.
Here, flat band systems \cite{Lieb:1989aa,Mielke:1991aa,Tasaki:1992aa,Misumi:2017aa,Rhim:2019aa}
have been attracting continuous interest, not only in magnetism
\cite{Lieb:1989aa,Mielke:1991aa,Tasaki:1992aa},
but also
the fractional QHE \cite{Sun:2011aa,Tang:2011aa,Neupert:2011aa,Yang:2012aa},
and 
superconductivity \cite{Kobayashi:2016aa,Tovmasyan:2016aa}, 
especially in twisted bilayer graphene 
\cite{Cao:2018aa,Fatemi:2018aa,Zou:2018aa,Rademaker:2018aa,Venderbos:2018aa,Ramires:2018aa,Koshino:2018aa,Peltonen:2018aa,Po:2018aa,Guo:2018aa,Ochi:2018aa,Lin:2018aa,Kennes:2018aa,Choi:2018aa,Qiao:2018aa,Hejazi:2019aa,Tarnopolsky:2019aa}, etc.
Thus, it may be interesting to investigate more generic flat bands with edge and/or corner states to 
seek the possibilities of wider topological phases of matter.
Such systems would offer a promising platform for studying the interplay among magnetism, superconductivity, and topological 
phenomena.

In this paper, we investigate a  model with inversion ($\cal P$) symmetry
as well as time-reversal ($\cal T$) symmetry, which is an extension of the two-dimensional 
Su-Schrieffer-Heeger (SSH) model
in Refs. \cite{Benalcazar:2017aa,Benalcazar:2017ab,Liu:2017aa}, including next-nearest neighbor (NNN) hoppings 
\cite{Misumi:2017aa}.
We first argue that topological invariants for a $\cal PT$-symmetric system
are quantized Berry phases
\cite{Yu:2015aa,Kim:2015aa,Huang:2016aa,Zhang:2016ab,Chan:2016aa,Pal:2018ab,Pal:2018aa,Li:2018aa,Ahn:2018aa,1806.11116,1808.05375,1810.02373,1810.04094,Ahn:2019aa,Ghatak_2019}
(or polarizations) 
which could be also topological invariants characterizing the HOTI.
Moreover, with fine-tuned parameters for the NNN hoppings, the model allows flat bands.
Therefore, we can realize  systems in which gapped edge states, corner states and flat bands coexist.
We show that the model introduced in this paper
has edge states and corner states on a flat-band like the breathing Kagome lattice model \cite{Ezawa:2018aa}.
The flat bands with corner states thus derived are near the critical point to flat Chern bands
in the sense that the band-crossing with a dispersive band and infinitesimal symmetry-breaking perturbations 
could turn these (trivial) flat bands into flat Chern bands. To show this, 
we introduce a locally fluctuating magnetic flux breaking both $\cal T$ and $\cal P$ symmetries
but preserving $\cal PT$ symmetry.
We show that by increasing a small $\cal PT$ symmetry-breaking perturbation,
quantized step-like polarizations are changed
into winding polarizations. 
This is achieved by infinitesimal $\cal P$-breaking perturbations with $\cal T$ invariance.

\section{$PT$ symmetry}

We first argue quantized Berry phases for $\cal PT$-symmetric systems
\cite{Yu:2015aa,Kim:2015aa,Huang:2016aa,Zhang:2016ab,Li:2018aa,1810.04094}.
Let $H({\bm k})$ be a Hamiltonian and $\psi({\bm k})$ be the  $n$-multiplet,
$H({\bm k})\psi({\bm k})=\psi({\bm k}){\cal E}({\bm k})$.
$\cal PT$ symmetry is described by
\begin{alignat}1
	PTH({\bm k})(PT)^{-1}=H({\bm k}),
	\label{Sym}
\end{alignat}
where $T=K$ stands for the complex conjugation, and $(PT)^2=1$.
This symmetry  imposes the constraint on the wave function such that $\psi({\bm k})=\left[PT\psi({\bm k})\right]U({\bm k})=P\psi^*({\bm k})U({\bm k})$,
where $U({\bm k})$ is a  certain $n\times n$ unitary matrix.
Let $A_\mu({\bm k})\equiv \tr \psi^\dagger({\bm k})\partial_{\mu}\psi({\bm k})$ be the U(1) Berry connection, where
$\partial_\mu=\partial_{k_\mu}$ is the partial derivative with respect to the momentum $k_\mu$.
Then, the constraint on the wave function yields the identity,
\begin{alignat}1
	A_\mu({\bm k})=-A_\mu({\bm k})+\tr U^\dagger({\bm k})\partial_\mu U({\bm k}).
	\label{BerCon}
\end{alignat}
This implies that the Berry connection is a pure gauge, and hence,
the polarization defined by
\begin{alignat}1
	p_x(k_y)\equiv\frac{1}{2\pi i}\int_{-\pi}^\pi A_x({\bm k})dk_x,
\end{alignat}
is quantized such that
\begin{alignat}1
	p_x(k_y)
	=\frac{n_x(k_y)}{2} \quad {\rm mod } ~1,
\end{alignat}
where $n_x(k_y)\equiv \frac{1}{2\pi i}\int \tr U^\dagger({\bm k})\partial_x U({\bm k}) dk_x$ is the winding number
generically dependent on $k_y$. Likewise, the polarization $p_y(k_x)$ is quantized as $p_y(k_x)=\frac{n_y(k_x)}{2}$ mod $1$.
If the multiplet under consideration is
isolated from other bands  by finite direct gaps over the whole Brillouin zone,
the integers $n_\mu$ ($\mu=x,y$)  should be constant, and
the set of polarizations $(p_x,p_y)$ can serve as the topological invariant for $\cal PT$-symmetric systems.
It follows from Eq. (\ref{BerCon}) that for isolated multiplet bands  the Berry curvature vanishes
and the multiplet has zero first Chern number.

Let us focus our attention next on one specific band among the multiplets.
The band degeneracies of $\cal PT$-symmetric systems have codimension 2, implying that the
2D systems have at most Dirac-like point nodes.
With such point nodes, the polarizations $(p_x(k_y),p_y(k_x))$ of the specific band
have jump discontinuities at those nodes.
These discontinuities give rise to $\delta$-function like singularities to Berry curvature.
As far as $\cal PT$ symmetry is preserved,
the charge of monopoles at point nodes is indistinguishable, 
because 
$p=1/2=-1/2$ modulo 1.
However,
infinitesimal $\cal PT$ symmetry breaking perturbations enables us to observe the charge. If the total monopole charge
behind the multiplet under consideration is finite, it yields a finite Chern number if $\cal PT$ symmetry is broken.
Therefore, flat bands in $\cal PT$-symmetric systems could be converted into flat Chern bands.
Interestingly, this can be achieved by $\cal P$-symmetry-breaking  (but $\cal T$-symmetry-preserving)
perturbations.

\begin{figure}[htb]
	\begin{center}
		\includegraphics[scale=0.5]{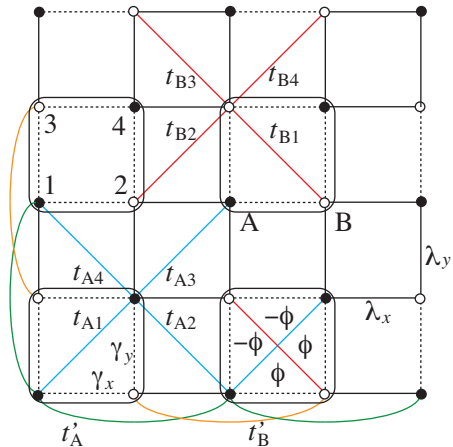}
		\caption{
			Schematic illustration of the model defined in Eq. (\ref{HamEle}).
			$\gamma_\mu$ and $\lambda_\mu$ ($\mu=x,y$) are nearest neighbor hopping, 
			denoting the SSH-like bond-alternating hoppings,
			whereas $t_{aj}$ and $t'_a$ ($a=$A,B and $j=1,\cdots,4$) are next-nearest neighbor hoppings necessary  for flat bands.
			In a lower unit cell, we denote  local flux $\phi$  which  breaks $\cal T$ and $\cal P$ symmetries but
			preserves $\cal PT$ symmetry.
		}
		\label{f:fig1}
	\end{center}
\end{figure}

\section{model}

We now study the 2D version of the SSH model (Fig. \ref{f:fig1}) including next-nearest neighbor hoppings,
$H({\bm k})=H_{\rm SSH}({\bm k})+H_{\rm NNN}({\bm k})$,
where $H_{\rm SSH}$ is the 2D generalization of the SSH model introduced in
\cite{Benalcazar:2017aa,Benalcazar:2017ab,Liu:2017aa}, and $H_{\rm NNN}$ is the next-nearest neighbor hopping terms
necessary for flat bands  \cite{Misumi:2017aa}. They are given by
\begin{alignat}1
	 & H_{\rm SSH}({\bm k})=
	\begin{pmatrix}
		0          & z_{x}^*(k_x) & z_{y}^*(k_y) & 0            \\
		z_{x}(k_x) & 0            & 0            & z_{y}^*(k_y) \\
		z_{y}(k_y) & 0            & 0            & z_{x}^*(k_x) \\
		0          & z_{y}(k_y)   & z_{x}(k_x)   & 0
	\end{pmatrix},
	\nonumber                \\
	 & H_{\rm NNN}({\bm k})=
	\begin{pmatrix}
		d_{\rm A}({\bm k}) & 0                  & 0                    & z_{\rm A}^*({\bm k}) \\
		0                  & d_{\rm B}({\bm k}) & z_{\rm B}^*({\bm k}) & 0                    \\
		0                  & z_{\rm B}({\bm k}) & d_{\rm B}({\bm k})   & 0                    \\
		z_{\rm A}({\bm k}) & 0                  & 0                    & d_{\rm A}({\bm k})
	\end{pmatrix},
	\label{HamEle}
\end{alignat}
where $z_{\mu}(k_\mu)\equiv \gamma_{\mu}+\lambda_\mu e^{ik_\mu}$  ($\mu=x,y$),
$z_{\rm A}({\bm k})=t_{\rm A1}+t_{\rm A2}e^{ik_x}+t_{\rm A4}e^{ik_y}+t_{\rm A3}e^{i(k_x+k_y)}$,
$z_{\rm B}({\bm k})=t_{\rm B1}+t_{\rm B2}e^{-ik_x}+t_{\rm B4}e^{ik_y}+t_{\rm B3}e^{i(-k_x+k_y)}$, and
$d_a({\bm k})=2t_a'(\cos k_x+\cos k_y)$ ($a=$A,B).
For the time being, we assume that
all the hopping parameters are real. Then,  this model has $\cal T$ symmetry and
$\cal P$ symmetry separately, where
$\cal P$ symmetry is implemented  by $P=\sigma_1\otimes\sigma_1$.
Moreover, when $\gamma_{x}=\gamma_{y}$, $\lambda_x=\lambda_y$, $t_{{\rm A}j}=t_{{\rm B}j}$ ($j=1,\cdots,4$),
and $t'_{\rm A}=t'_{\rm B}$, $H({\bm k})$ has C$_4$ symmetry.
This model interpolates the 2D SSH model \cite{Benalcazar:2017ab,Benalcazar:2017aa,Liu:2017aa}
and the flat band model \cite{Misumi:2017aa}: The latter model is  reproduced
when $\gamma_\mu=\lambda_\mu$ for $\mu=x,y$ and $t_{a}'=t_{aj}/2$ for $a=$A,B and $j=1,\cdots,4$.
For simplicity, we set $\lambda_x=\lambda_y=\lambda$  and
$\gamma_x=\gamma_y\equiv\gamma$ in this paper.
Thus, we assume C$_4$ symmetry in the basic $H_{\rm SSH}$ part.
Then, based on the corner states of exactly solvable $H_{\rm SSH}$, 
we will investigate the effect of $H_{\rm NNN}$, which is responsible for flat bands.

\subsection{Solvable decoupled model}

Let us first review the model without $H_{\rm NNN}$  studied in Ref. \cite{Liu:2017aa}.
The SSH part is exactly solvable, since
$H_{\rm SSH}({\bm k})=h_{\rm SSH}(k_x)\otimes\1+\1\otimes h_{\rm SSH}(k_y)$,
where $h_{\rm SSH}$ is the conventional 1D SSH model  \cite{Su:1979aa}.
Let $\psi_n(k)$ be the eigenstate of $h_{\rm SSH}(k)$ such that
$h_{\rm SSH}(k)\psi_n(k)=\varepsilon_n(k)\psi_n(k)$, where $n=\pm$ stands for the positive and negative bands and
$\varepsilon_n(k)$ ($\varepsilon_-(k)=-\varepsilon_+(k)$) is the energy eigenvalue.
Cleary, the eigenstates and corresponding eigenvalues of $H_{\rm SSH}({\bm k})$ are given by $\psi_{nm}({\bm k})=\psi_n(k_x)\otimes\psi_m(k_y)$ and
$\varepsilon_{nm}({\bm k})=\varepsilon_n(k_x)+\varepsilon_m(k_y)$, respectively.
It follows that the present 2D model has four bands, one negative $\varepsilon_{--}$, one positive $\varepsilon_{++}$, and
doubly-degenerate nearly zero-energy bands $\varepsilon_{+-}$ and $\varepsilon_{-+}$.
Each of these four bands has  the polarizations $(p_x,p_y)=(1/2,1/2)$ for $|\gamma/\lambda|<1$ and $(0,0)$ otherwise.
Therefore, at $1/4$- and $3/4$-filling, the polarizations of the ground state are nontrivial when
$|\gamma/\lambda|<1$.

\begin{figure}[htb]
	\begin{center}
		\begin{tabular}{cc}
			\includegraphics[scale=0.32]{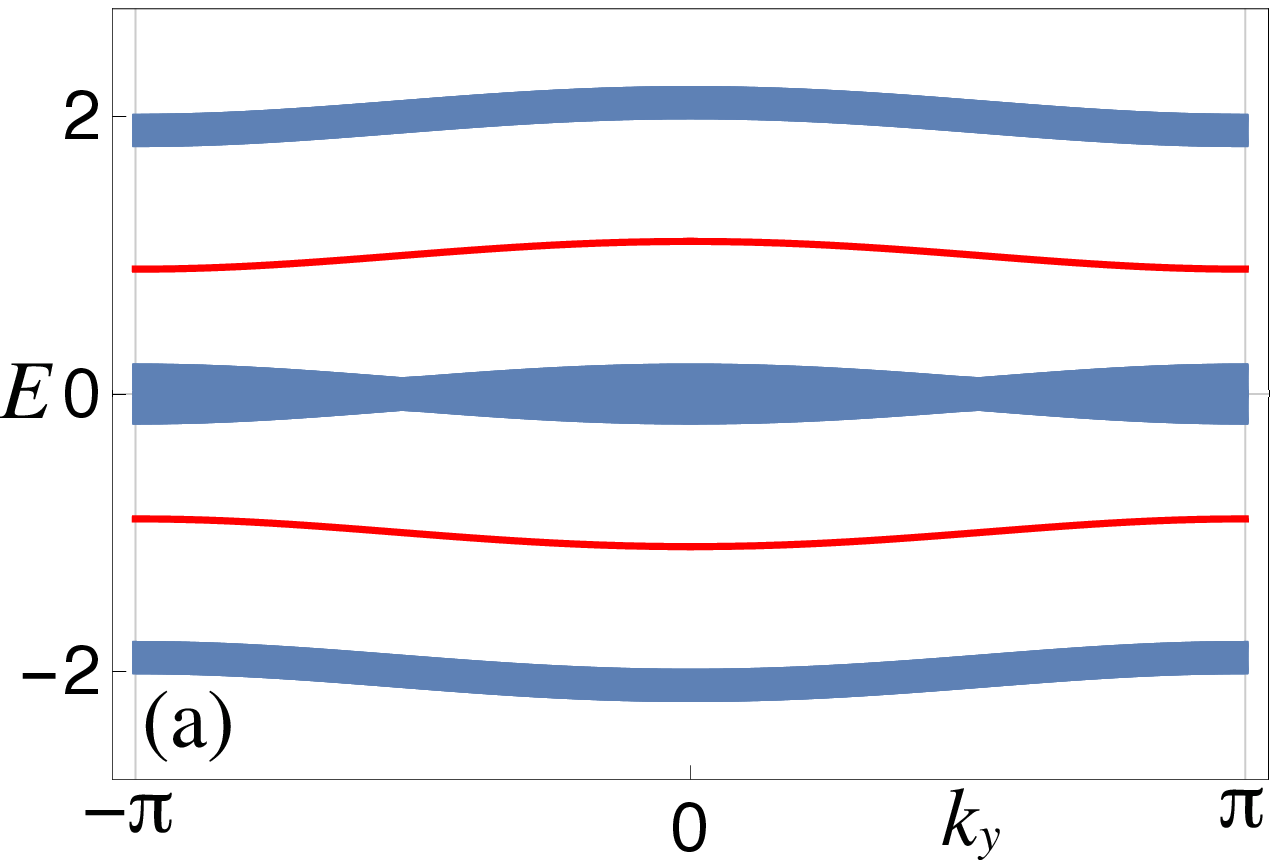} &
			\includegraphics[scale=0.33]{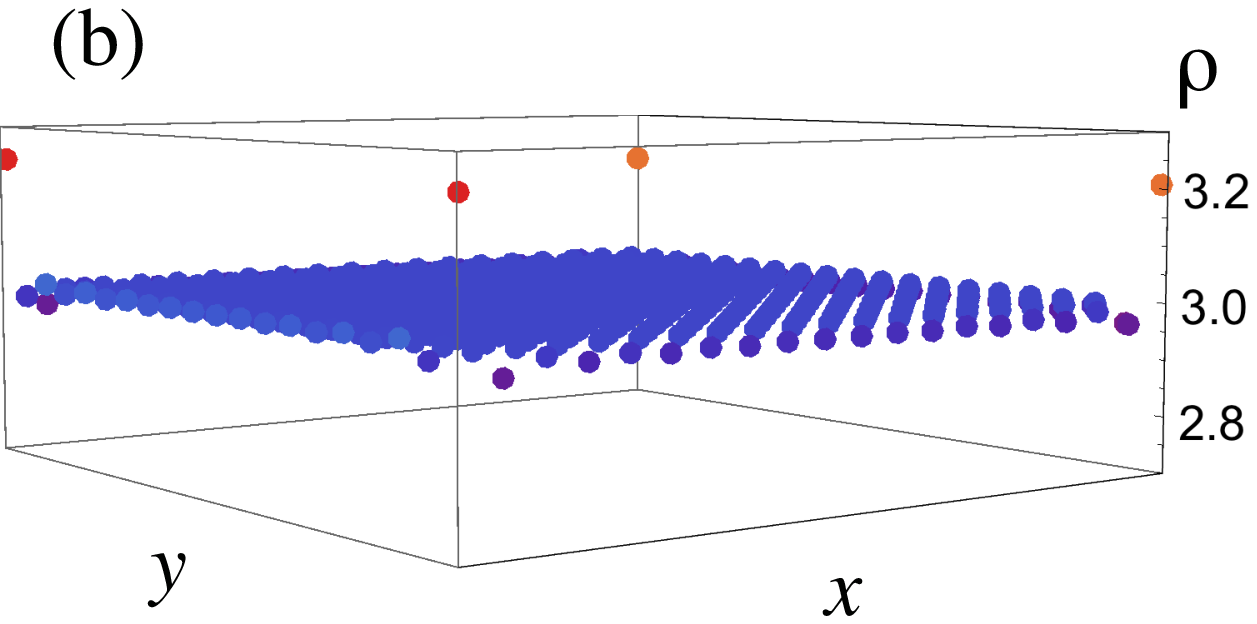}
		\end{tabular}
		\caption{
			(a) Energy spectrum of the cylindrical system with $\lambda=1$ and $\gamma=0.1$,
			in which edge states denoted by red lines can be observed.
			(b) Particle density at $3/4$-filling.}
		\label{f:fig2}
	\end{center}
\end{figure}


Such a decoupling property also holds for systems with boundaries.
For the cylindrical system with an open boundary condition toward the $x$ direction,
the wave function of the 2D system is the tensor product of the wave function of the open SSH chain with respect to $x$ and
the wave function of the periodic SSH chain with respect to $y$, and
the total spectrum  is just given by the sum of corresponding energies.
Therefore, it is obvious that  in the case  $|\gamma/\lambda|<1$,
the cylindrical system has gapped edge states,
as shown in Fig. \ref{f:fig2} (a),
 which are a combination of
the Bloch states in the $y$ direction
and the edge states in the $x$ direction.
These edge states are themselves the 1D SSH states associated with $\psi_n(k_y)$, and their polarization
is clearly $p_y=1/2$.
Therefore, for a full open system with four corners, the model shows
corner states
exactly at zero energy, even though they are embedded in the bulk spectrum.
It thus turns out that the corner states are protected by the nontrivial polarizations $(p_x,p_y)$.
In addition to the corner states, the model also has edge states, and
since the Fermi energy lies within these edge states
at $3/4$- ($1/4$-) filling,
there appear corner particle (hole) states affected by edge states
\cite{Benalcazar:2017aa,Benalcazar:2017ab,Liu:2017aa,Schindler:2018ab}.
Figure \ref{f:fig2} (b) shows the particle density at $3/4$-filling for
the system in Fig. \ref{f:fig2} (a) with full open boundary conditions imposed.
One can see sharp peaks  at four corners as well as the edge state contribution around the boundaries
deviated slightly  from the average density of the bulk.

\begin{figure}[htb]
	\begin{center}
		\begin{tabular}{cc}
			\includegraphics[scale=0.3]{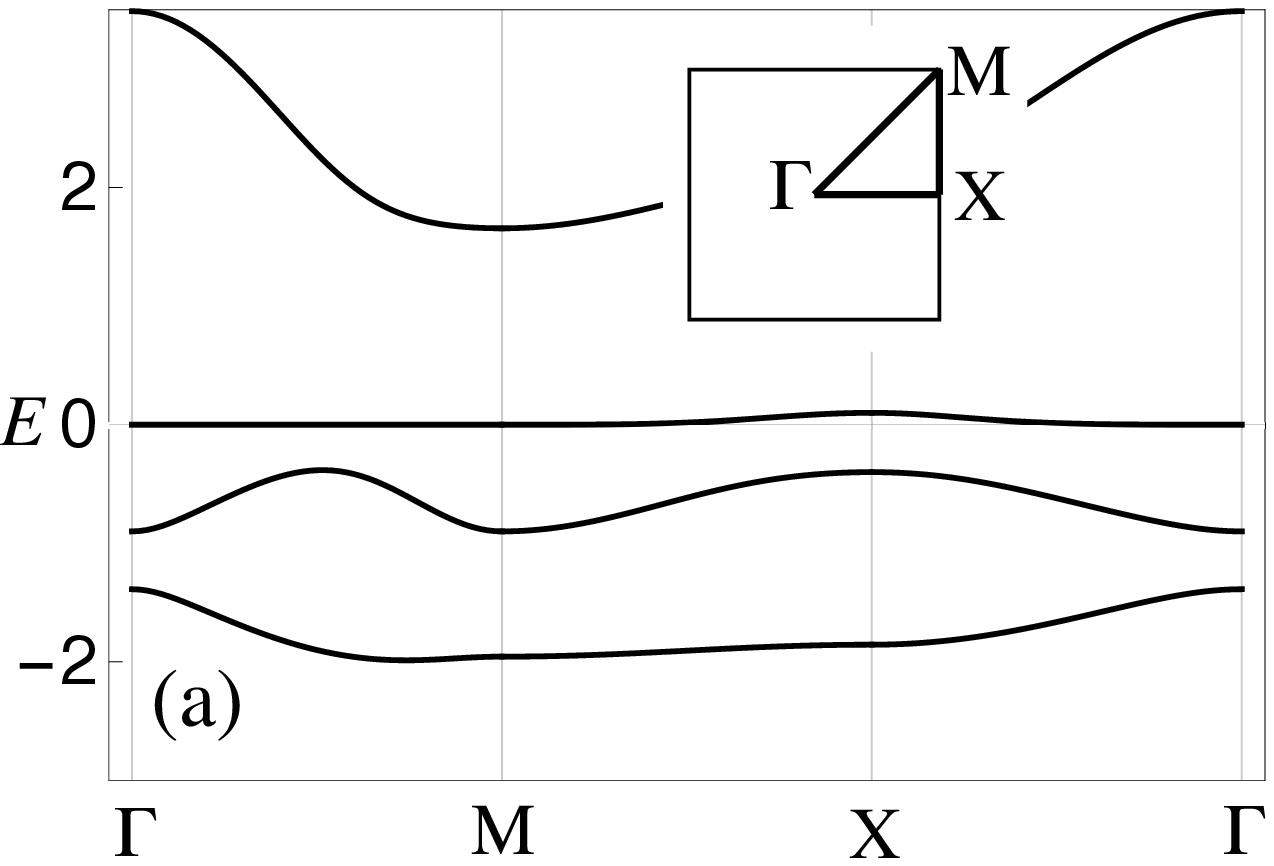}
			 &
			\includegraphics[scale=0.3]{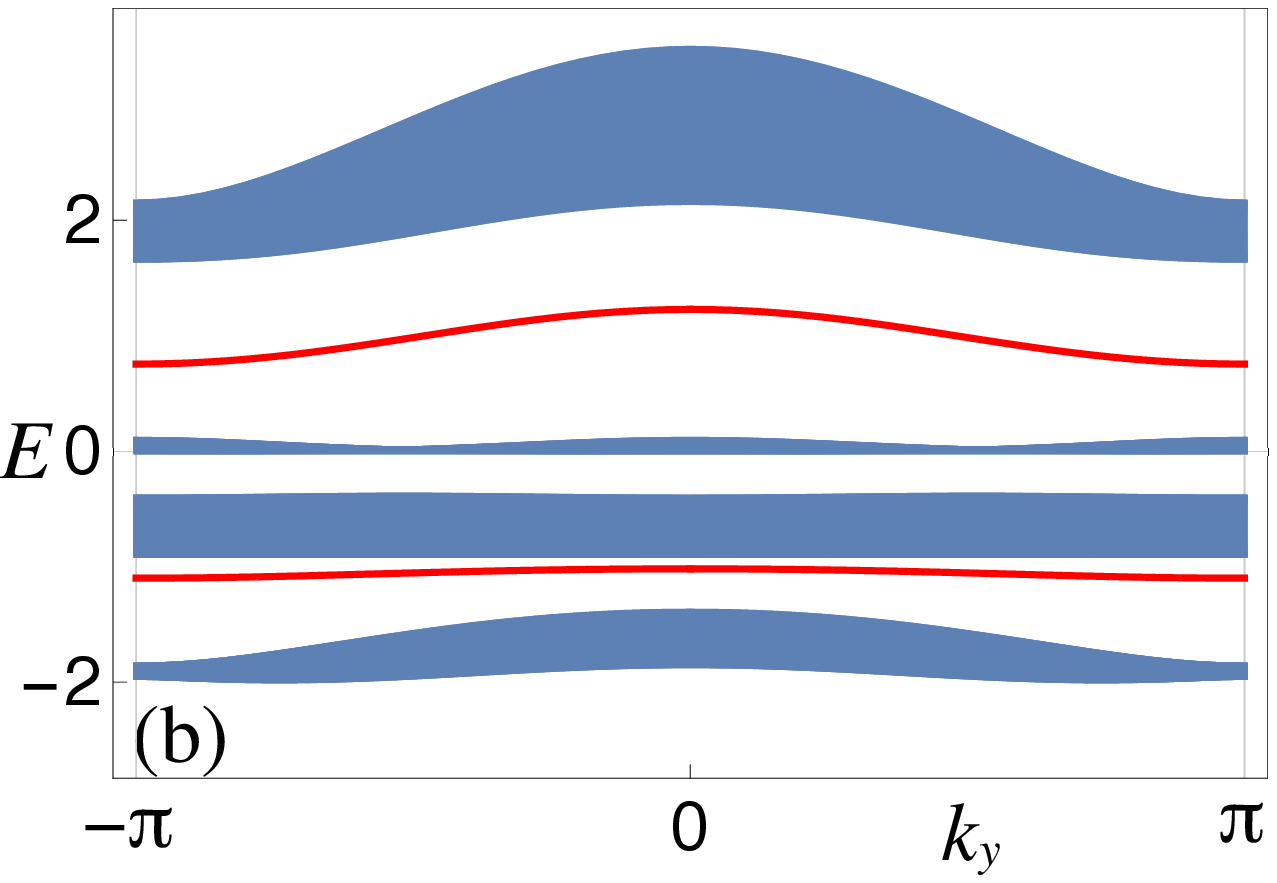}
			\\
			\includegraphics[scale=0.3]{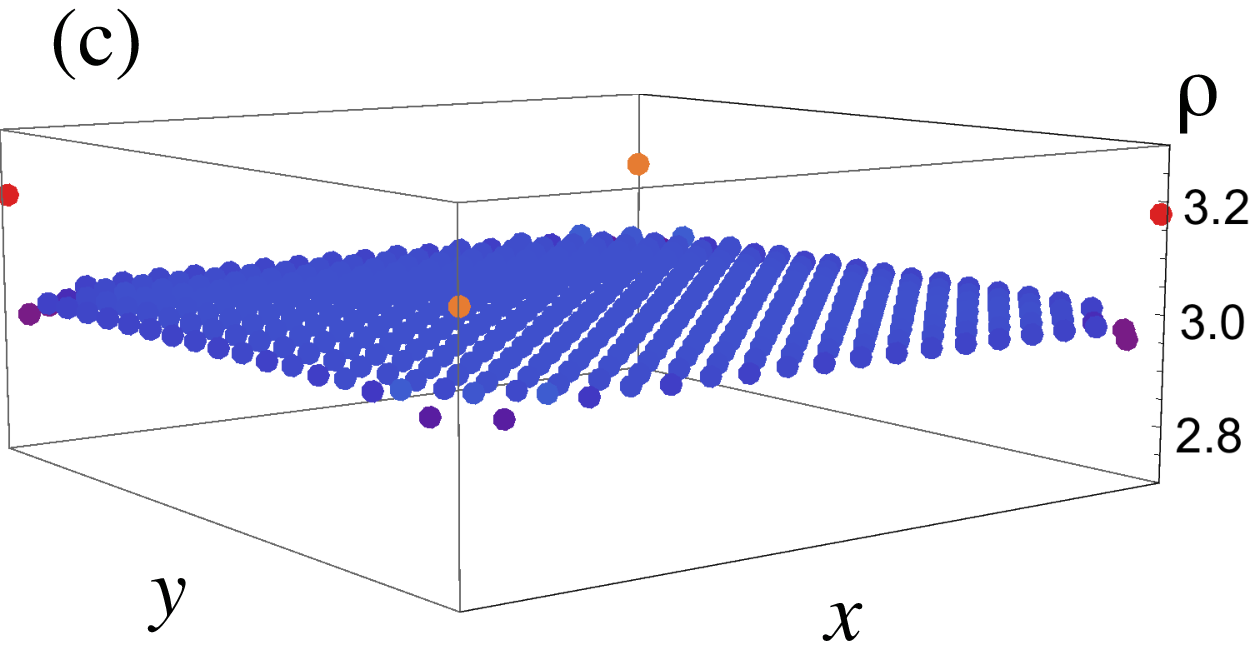}
			 &
			\includegraphics[scale=0.3]{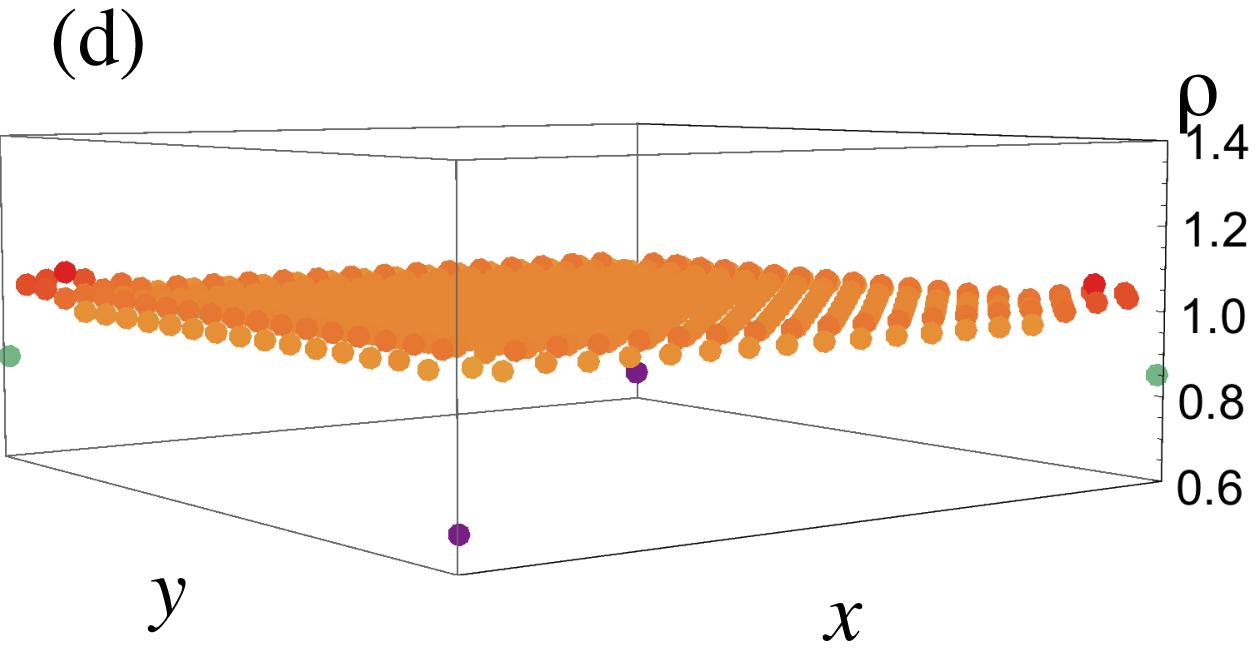}
		\end{tabular}
		\caption{
			(a) Spectrum of the bulk system. The inset shows the Brillouin zone.
			(b) Spectrum of the system with open boundary conditions toward the $x$ direction.
			The edge states are highlighted by red lines.
			(c) and (d) Particle densities at 3/4- and 1/4-fillings, respectively.
			The parameters used are $\lambda=1$, $\gamma=0.1$,
			$t_{\rm A}=0.3$, and $\epsilon_{\rm A}=2$.
		}
		\label{f:fig3}
	\end{center}
\end{figure}

\subsection{Next-nearest neighbor hopping term and flat band}

Let us next include next-nearest neighbor hoppings $H_{\rm NNN}$ to investigate
flat bands with corner states.
In the case of $\lambda=\gamma$, $t_{{\rm A}j}=t'_{\rm A}/2$,
and $t_{{\rm B}j}=t'_{\rm B}=0$, this model coincides with the model
proposed in \cite{Misumi:2017aa},
implying that the Hamiltonian (\ref{HamEle}) has a complete flat band
separated from a dispersive band by a finite gap.
On the other hand, when $\lambda=0$ or $\gamma=0$ as well as $t_{{\rm B}j}=t'_{\rm B}=0$,
one can prove that the model also
has an exact flat band at zero energy for any other parameters $t_{{\rm A}j}$ and $t'_{\rm A}$.
For simplicity, we restrict our model to
\begin{alignat}1
	 & t_{{\rm A}j}=2t'_{\rm A}=t_{\rm A}\quad (j=1,2,4),\quad t_{{\rm A}3}=\epsilon_{\rm A} t_{\rm A}
	\nonumber                                                                                          \\
	 & t_{{\rm B}j}=t'_{\rm B}=0.
	\label{Par}
\end{alignat}
Figure \ref{f:fig3} (a) shows the spectrum of the model with small $\gamma=0.1$.
The exact flat band when $\gamma=0$ becomes slightly dispersive but is still an almost flat band.
All four bands have the nontrivial polarizations $(1/2,1/2)$.
Therefore, the system with the open boundary condition toward, e.g., the $x$ direction should have the edge states
within the first gap and the third gap guaranteed by $p_x=1/2$, as shown in Fig. \ref{f:fig3} (b).
Moreover, the nontrivial polarization $p_y=1/2$ suggests that these edge states would form corner states around zero energy
if we  further impose the open boundary condition toward the $y$ direction.

Indeed, the spectrum in Fig. \ref{f:fig3} (b) can be obtained by an adiabatic deformation of the spectrum
in Fig. \ref{f:fig2} (c), implying that the edge states in both figures have the same topological property.
Namely, those in Fig. \ref{f:fig3} (b) have also polarization $p_y=1/2$, and they are 1D topological insulators.
Thus, for a full open system, corner states should emerge in between those edge states, probably around zero energy.
We show in Figs. \ref{f:fig3} (c) and (d) the particle density for a system with full open boundary condition
at 3/4- and 1/4-filling, respectively.
One can observe particle-like and hole-like densities at four corners, which are due to fully occupied and fully unoccupied
corner states.  A slight effect of  the edge states can also be seen,
since the Fermi energy is just on the edge states.

\begin{figure}[htb]
	\begin{center}
		\begin{tabular}{cc}
			\includegraphics[scale=0.3]{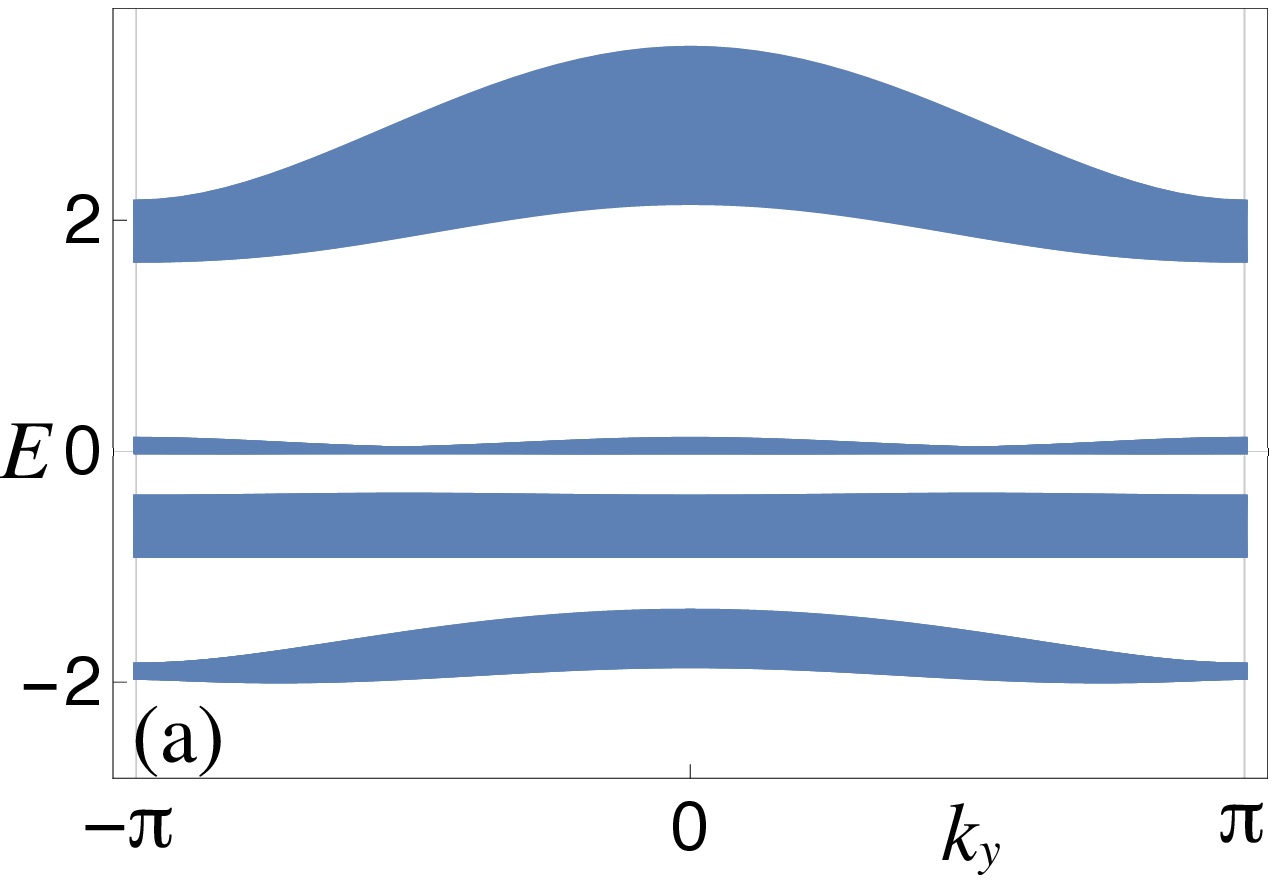}
			 &
			\includegraphics[scale=0.3]{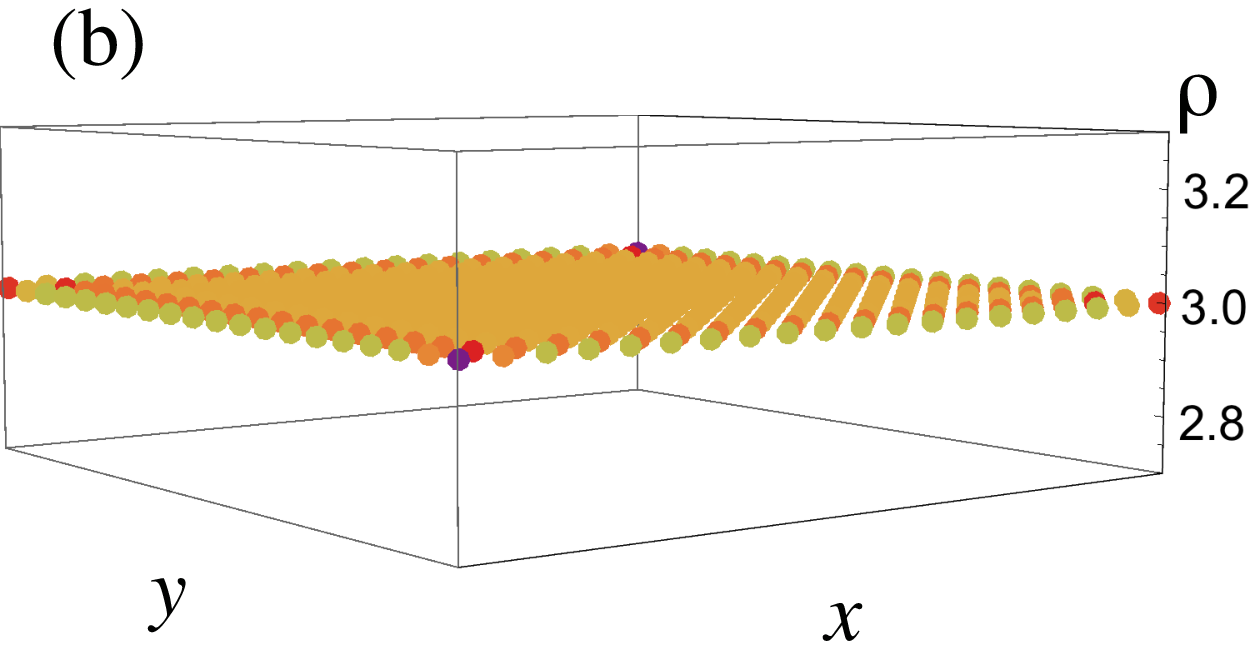}
		\end{tabular}
		\caption{
			(a) Spectrum of the system with open boundary conditions toward the $x$ direction.
						(b) Particle density at 3/4-filling.
			The parameters used are replaced by $\lambda\leftrightarrow\gamma$,
			$t_{{\rm A}1}\leftrightarrow t_{{\rm A}3}$ in Fig. \ref{f:fig3}.
		}
		\label{f:fig4}
	\end{center}
\end{figure}

These edge and corner states are in sharp contrast to the trivial case,  particularly,
for the  system with replaced parameters $\lambda\leftrightarrow\gamma$
(and also $t_{{\rm A}1}\leftrightarrow t_{{\rm A}3}$) which has the same bulk spectrum by definition.
In this system, the polarizations of each  band is (0,0), and therefore, there are no edge states for a cylindrical system,
as shown in Fig. \ref{f:fig4} (a). Correspondingly, the corner states vanish as in Fig. \ref{f:fig4} (b) for a full open system.

\begin{figure}[htb]
	\begin{center}
		\begin{tabular}{cc}
			\includegraphics[scale=0.3]{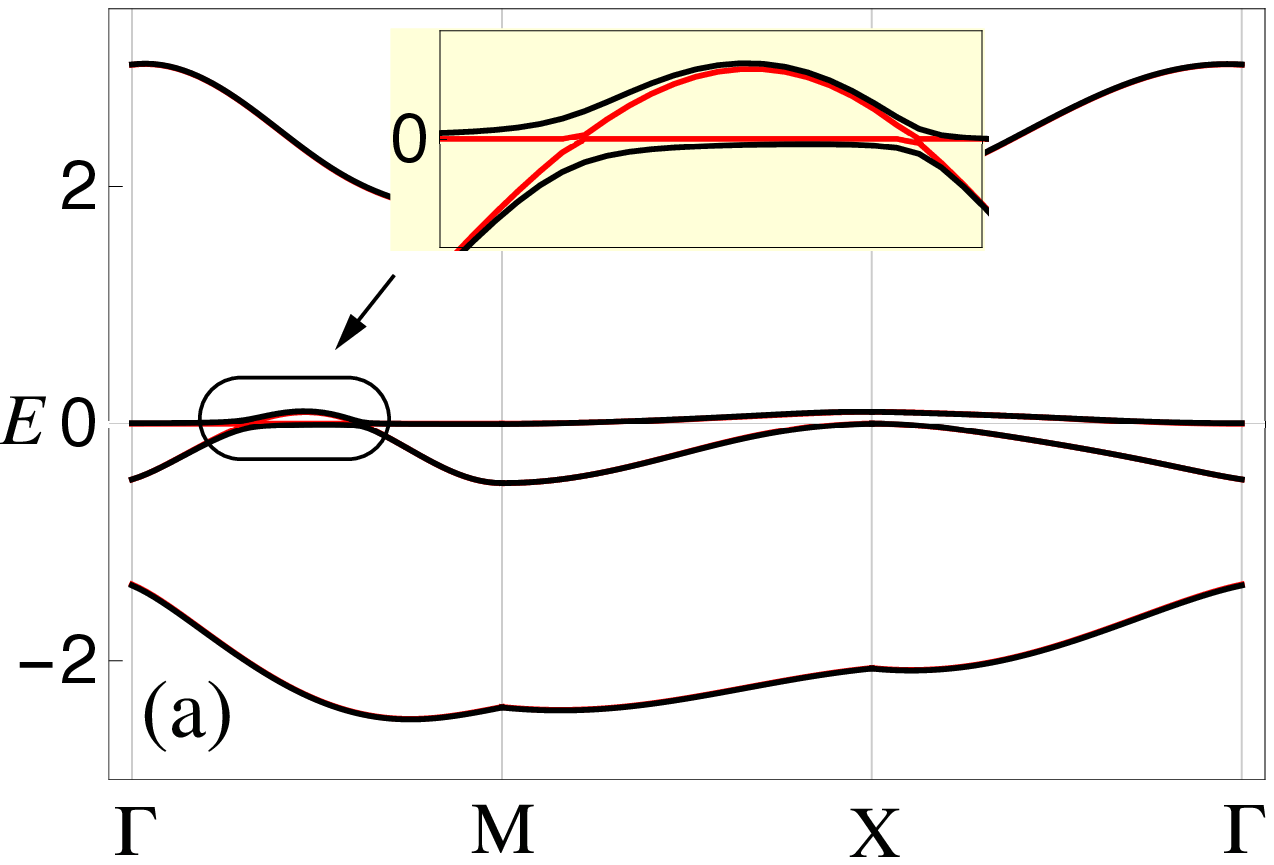}
			 &
			\includegraphics[scale=0.33]{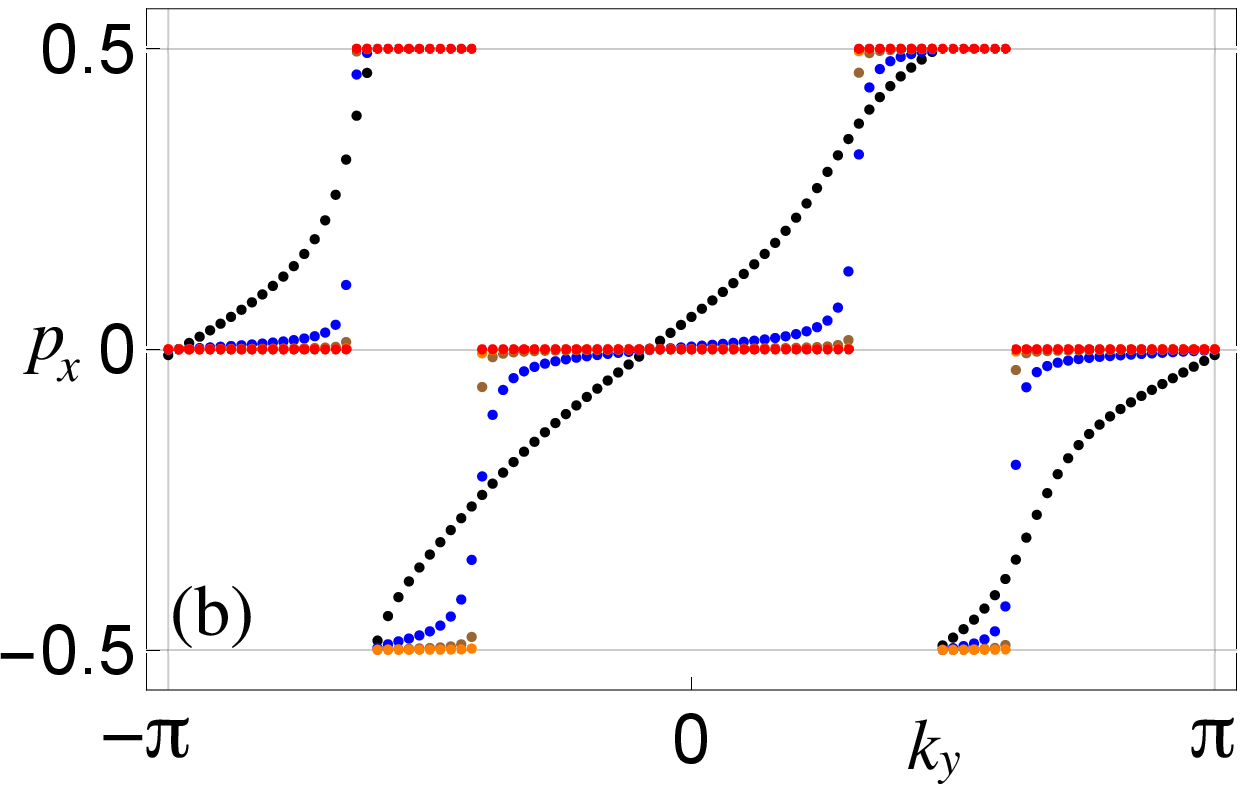}
			\\
			\includegraphics[scale=0.3]{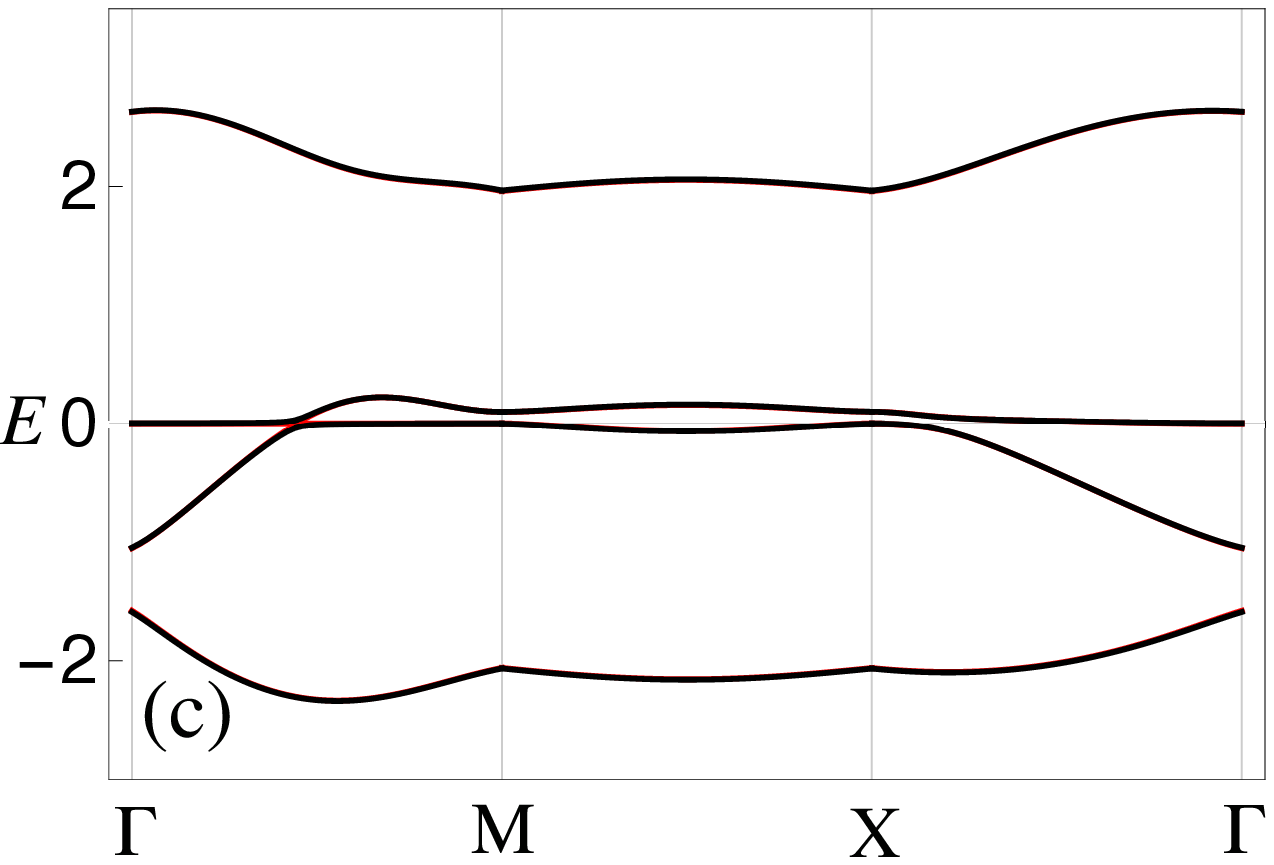}
			 &
			\includegraphics[scale=0.33]{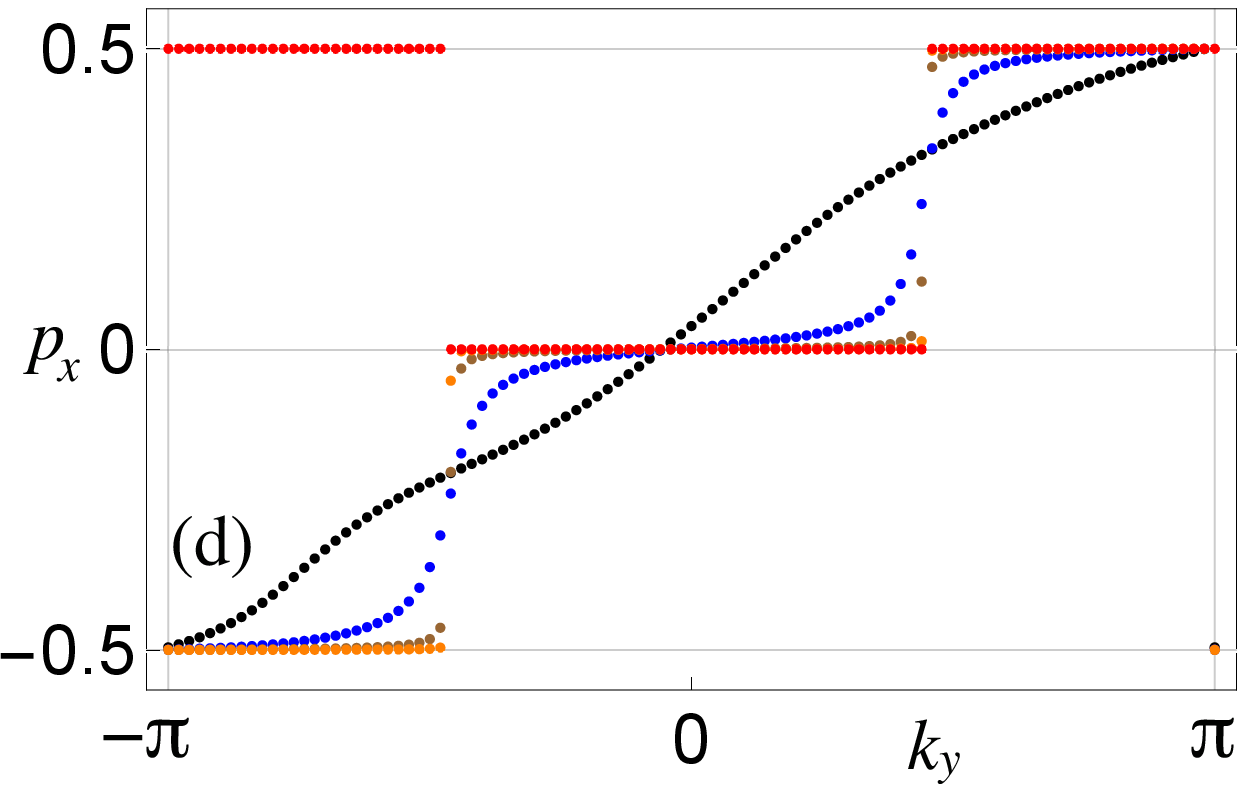}
		\end{tabular}
		\caption{
		(a) Spectrum with a flat band with a symmetry breaking potential. The red (black)  lines show the spectrum with
		$v_{\rm sb}=0$ ($0.1$). Inset shows the gap formation near the point nodes.
		The other parameters are
		$\gamma=0.1 e^{i\phi}$ with $\phi=\pi/2$, $t_{{\rm A}}=0.3$, and $\epsilon_{{\rm A}}=2/3$.
		(b) Polarization $p_x(k_y)$ of the flat (third) band. The red, orange, brown, blue, black dots are the cases with
		$v_{\rm sb}=0$, $10^{-4}$, $10^{-3}$,$10^{-2}$, $10^{-1}$, respectively.
		(c) Spectrum  and (d) the polarization $p_x(k_y)$ for the same system as (a) and (b) except for the
		$t'_{\rm A}=0$.
		}
		\label{f:fig5}
	\end{center}
\end{figure}

\subsection{$\cal PT$ symmetry breaking and flat Chern band}

So far we have discussed the edge and corner states
of the nearly flat band with nontrivial polarizations $(p_x,p_y)=(1/2,1/2)$.
The quantization of these polarizations is due to $\cal PT$ symmetry, which also ensures
vanishing Berry curvatures and Chern numbers.
We next consider the possibility of converting such a flat band into a flat Chern band.
First, band inversions with dispersive bands may be necessary.
For the present model, this can be controlled by the parameter $\epsilon_{\rm A}$ in Eq. (\ref{Par}).
Second, ${\cal T}$ symmetry should be broken.
To this end, let us introduce locally  fluctuating magnetic flux with zero mean \cite{Haldane:1988aa},
as denoted in Fig. \ref{f:fig1}.
It can be incorporated simply by replacing $\gamma_\mu$ with $\gamma_\mu\rightarrow \gamma_\mu e^{i\phi}$
for $\mu=x,y$.
Such flux  $\phi$ breaks not only $\cal T$ symmetry but also $\cal P$ symmetry.
Nevertheless,  $\cal PT$ symmetry is preserved, implying that the flat band is still a trivial Chern band.
In Fig. \ref{f:fig5}, we show the spectrum (red lines in (a)) and the polarization $p_x(k_y)$  (red dots in (b))
of the flat band for systems with $\cal PT$ symmetry only, including finite local flux $\phi$.
The quantized polarization $p_x(k_y)$ of the flat band due to $\cal PT$ symmetry
has several jump discontinuities. These are a hallmark of  Dirac-like level-crossings, which indeed occur
as point nodes because of the codimension 2 of $\cal PT$-symmetric systems.
These point nodes give rise to singular Berry curvature.
Finally, if $\cal PT$ symmetry-breaking perturbations are introduced, gaps are formed at these point nodes, and
discontinuities of the polarization could become smoothly winding between $k_\mu=-\pi$ and $\pi$.
The winding number of the polarization is merely the Chern number, and thus, a trivial flat band would be
converted into a flat Chern band.
To  show this,
we introduce $\cal P$-symmetry breaking  potentials to the Hamiltonian 
\begin{alignat}1
	H_{\rm on}=v_{\rm sb}\sigma_3\otimes\1+v'_{\rm sb}\1\otimes\sigma_3.
	\label{Per}
\end{alignat}
These real onsite potentials of course preserve $\cal T$ symmetry, but they break $\cal PT$ symmetry.
Below, we take only the $v_{\rm sb}$ term into account in Eq. (\ref{Per}), setting $v_{\rm sb}'=0$, for simplicity.
In Fig. \ref{f:fig5} (a), we show the spectrum of the system with a small $\cal P$ symmetry breaking potential
(black lines).
One can observe that small gaps are indeed formed at the point nodes. 
As can be seen from the polarization $p_x(k_y)$ of the flat band denoted by black points in Fig. \ref{f:fig5} (b),
the winding number of $p_x$ is 2, impling the Chern number $-2$. The direct computation of Chern numbers
\cite{FHS05} also supports this result.
In this figure, one also sees how the step-like function of the polarization for a $\cal PT$ symmetric system
changes into a function with a winding number, when one varies the symmetry-breaking potential.
Basically, for infinitesimal perturbations, the flat band becomes a flat Chern band.
In this sense, the present $\cal PT$ symmetric flat band is just on the critical point from a trivial band to a Chern band.
Another example is shown in Fig. \ref{f:fig5} (c) and (d) which is the same model as in (a) and (b)
except for $t'_{\rm A}=0$. The winding number of the polarization implies the Chern number $-1$.

\section{Summary and discussions}

In summary, we have studied a $\cal PT$ symmetric system having a flat band together
with  edge and corner states
due to nontrivial polarizations.
We have argued that the flat band of the model could be changed into a flat Chern band
by symmetry-breaking  perturbations.
For flat band superconductors, the intersection of flat bands and dispersive bands 
is important \cite{Kobayashi:2016aa,Tovmasyan:2016aa,Misumi:2017aa}.
Thus, it may be interesting to consider the possibility of flat band superconductors with edge and/or corner states,
especially paying attention to the role played by those states.
As argued in \cite{Benalcazar:2019aa}, C$_n$ group allows various fractional charges at corners.
It may thus be interesting to consider the interplay between flat bands and edge and/or corner states in systems with
	more generic point group symmetry \cite{Benalcazar:2019aa,Fang:2012aa}.
	We also note that without C$_4$ symmetry, $H_{\rm SSH}$ Hamiltonian has unusual corner states inherent to 2D 
	nature of the model \cite{Li:2018ab}. It may be quite interesting to study such a model including $H_{\rm NNN}$.


We would like to thank  H. Aoki for fruitful discussions.
This work was supported in part by Grants-in-Aid for Scientific Research Numbers 17K05563 and 17H06138
from the Japan Society for the Promotion of Science.


\begin{thebibliography}{81}
\expandafter\ifx\csname natexlab\endcsname\relax\def\natexlab#1{#1}\fi
\expandafter\ifx\csname bibnamefont\endcsname\relax
  \def\bibnamefont#1{#1}\fi
\expandafter\ifx\csname bibfnamefont\endcsname\relax
  \def\bibfnamefont#1{#1}\fi
\expandafter\ifx\csname citenamefont\endcsname\relax
  \def\citenamefont#1{#1}\fi
\expandafter\ifx\csname url\endcsname\relax
  \def\url#1{\texttt{#1}}\fi
\expandafter\ifx\csname urlprefix\endcsname\relax\def\urlprefix{URL }\fi
\providecommand{\bibinfo}[2]{#2}
\providecommand{\eprint}[2][]{\url{#2}}

\bibitem[{\citenamefont{Hatsugai}(1993)}]{Hatsugai:1993fk}
\bibinfo{author}{\bibfnamefont{Y.}~\bibnamefont{Hatsugai}},
  \bibinfo{journal}{Physical Review Letters} \textbf{\bibinfo{volume}{71}},
  \bibinfo{pages}{3697} (\bibinfo{year}{1993}).

\bibitem[{\citenamefont{Thouless et~al.}(1982)\citenamefont{Thouless, Kohmoto,
  Nightingale, and den Nijs}}]{Thouless:1982uq}
\bibinfo{author}{\bibfnamefont{D.~J.} \bibnamefont{Thouless}},
  \bibinfo{author}{\bibfnamefont{M.}~\bibnamefont{Kohmoto}},
  \bibinfo{author}{\bibfnamefont{M.~P.} \bibnamefont{Nightingale}},
  \bibnamefont{and} \bibinfo{author}{\bibfnamefont{M.}~\bibnamefont{den Nijs}},
  \bibinfo{journal}{Physical Review Letters} \textbf{\bibinfo{volume}{49}},
  \bibinfo{pages}{405} (\bibinfo{year}{1982}).

\bibitem[{\citenamefont{Kohmoto}(1985)}]{kohmoto:85}
\bibinfo{author}{\bibfnamefont{M.}~\bibnamefont{Kohmoto}},
  \bibinfo{journal}{Annals of Physics} \textbf{\bibinfo{volume}{160}},
  \bibinfo{pages}{343} (\bibinfo{year}{1985}).

\bibitem[{\citenamefont{Kane and Mele}(2005)}]{Kane:2005aa}
\bibinfo{author}{\bibfnamefont{C.~L.} \bibnamefont{Kane}} \bibnamefont{and}
  \bibinfo{author}{\bibfnamefont{E.~J.} \bibnamefont{Mele}},
  \bibinfo{journal}{Physical Review Letters} \textbf{\bibinfo{volume}{95}},
  \bibinfo{pages}{146802} (\bibinfo{year}{2005}).

\bibitem[{\citenamefont{Qi et~al.}(2008)\citenamefont{Qi, Hughes, and
  Zhang}}]{Qi:2008aa}
\bibinfo{author}{\bibfnamefont{X.-L.} \bibnamefont{Qi}},
  \bibinfo{author}{\bibfnamefont{T.~L.} \bibnamefont{Hughes}},
  \bibnamefont{and} \bibinfo{author}{\bibfnamefont{S.-C.} \bibnamefont{Zhang}},
  \bibinfo{journal}{Physical Review B} \textbf{\bibinfo{volume}{78}},
  \bibinfo{pages}{195424} (\bibinfo{year}{2008}).

\bibitem[{\citenamefont{Schnyder et~al.}(2008)\citenamefont{Schnyder, Ryu,
  Furusaki, and Ludwig}}]{Schnyder:2008aa}
\bibinfo{author}{\bibfnamefont{A.~P.} \bibnamefont{Schnyder}},
  \bibinfo{author}{\bibfnamefont{S.}~\bibnamefont{Ryu}},
  \bibinfo{author}{\bibfnamefont{A.}~\bibnamefont{Furusaki}}, \bibnamefont{and}
  \bibinfo{author}{\bibfnamefont{A.~W.~W.} \bibnamefont{Ludwig}},
  \bibinfo{journal}{Physical Review B} \textbf{\bibinfo{volume}{78}},
  \bibinfo{pages}{195125} (\bibinfo{year}{2008}).

\bibitem[{\citenamefont{K{\"o}nig et~al.}(2008)\citenamefont{K{\"o}nig,
  Buhmann, Molenkamp, Hughes, Liu, Qi, and Zhang}}]{Konig:1977fk}
\bibinfo{author}{\bibfnamefont{M.}~\bibnamefont{K{\"o}nig}},
  \bibinfo{author}{\bibfnamefont{H.}~\bibnamefont{Buhmann}},
  \bibinfo{author}{\bibfnamefont{L.~W.} \bibnamefont{Molenkamp}},
  \bibinfo{author}{\bibfnamefont{T.}~\bibnamefont{Hughes}},
  \bibinfo{author}{\bibfnamefont{C.-X.} \bibnamefont{Liu}},
  \bibinfo{author}{\bibfnamefont{X.-L.} \bibnamefont{Qi}}, \bibnamefont{and}
  \bibinfo{author}{\bibfnamefont{S.-C.} \bibnamefont{Zhang}},
  \bibinfo{journal}{J. Phys. Soc. Jpn.} \textbf{\bibinfo{volume}{77}},
  \bibinfo{pages}{031007} (\bibinfo{year}{2008}).

\bibitem[{\citenamefont{Hasan and Kane}(2010)}]{Hasan:2010fk}
\bibinfo{author}{\bibfnamefont{M.~Z.} \bibnamefont{Hasan}} \bibnamefont{and}
  \bibinfo{author}{\bibfnamefont{C.~L.} \bibnamefont{Kane}},
  \bibinfo{journal}{Reviews of Modern Physics} \textbf{\bibinfo{volume}{82}},
  \bibinfo{pages}{3045} (\bibinfo{year}{2010}).

\bibitem[{\citenamefont{Qi and Zhang}(2011)}]{Qi:2011kx}
\bibinfo{author}{\bibfnamefont{X.-L.} \bibnamefont{Qi}} \bibnamefont{and}
  \bibinfo{author}{\bibfnamefont{S.-C.} \bibnamefont{Zhang}},
  \bibinfo{journal}{Reviews of Modern Physics} \textbf{\bibinfo{volume}{83}},
  \bibinfo{pages}{1057} (\bibinfo{year}{2011}).

\bibitem[{\citenamefont{Ando}(2013)}]{Ando:2013aa}
\bibinfo{author}{\bibfnamefont{Y.}~\bibnamefont{Ando}},
  \bibinfo{journal}{Journal of the Physical Society of Japan}
  \textbf{\bibinfo{volume}{82}}, \bibinfo{pages}{102001}
  (\bibinfo{year}{2013}).

\bibitem[{\citenamefont{Slager et~al.}(2015)\citenamefont{Slager, Rademaker,
  Zaanen, and Balents}}]{Slager:2015aa}
\bibinfo{author}{\bibfnamefont{R.-J.} \bibnamefont{Slager}},
  \bibinfo{author}{\bibfnamefont{L.}~\bibnamefont{Rademaker}},
  \bibinfo{author}{\bibfnamefont{J.}~\bibnamefont{Zaanen}}, \bibnamefont{and}
  \bibinfo{author}{\bibfnamefont{L.}~\bibnamefont{Balents}},
  \bibinfo{journal}{Physical Review B} \textbf{\bibinfo{volume}{92}},
  \bibinfo{pages}{085126} (\bibinfo{year}{2015}).

\bibitem[{\citenamefont{Benalcazar
  et~al.}(2017{\natexlab{a}})\citenamefont{Benalcazar, Bernevig, and
  Hughes}}]{Benalcazar:2017aa}
\bibinfo{author}{\bibfnamefont{W.~A.} \bibnamefont{Benalcazar}},
  \bibinfo{author}{\bibfnamefont{B.~A.} \bibnamefont{Bernevig}},
  \bibnamefont{and} \bibinfo{author}{\bibfnamefont{T.~L.}
  \bibnamefont{Hughes}}, \bibinfo{journal}{Physical Review B}
  \textbf{\bibinfo{volume}{96}}, \bibinfo{pages}{245115}
  (\bibinfo{year}{2017}{\natexlab{a}}).

\bibitem[{\citenamefont{Benalcazar
  et~al.}(2017{\natexlab{b}})\citenamefont{Benalcazar, Bernevig, and
  Hughes}}]{Benalcazar:2017ab}
\bibinfo{author}{\bibfnamefont{W.~A.} \bibnamefont{Benalcazar}},
  \bibinfo{author}{\bibfnamefont{B.~A.} \bibnamefont{Bernevig}},
  \bibnamefont{and} \bibinfo{author}{\bibfnamefont{T.~L.}
  \bibnamefont{Hughes}}, \bibinfo{journal}{Science}
  \textbf{\bibinfo{volume}{357}}, \bibinfo{pages}{61}
  (\bibinfo{year}{2017}{\natexlab{b}}).

\bibitem[{\citenamefont{Liu and Wakabayashi}(2017)}]{Liu:2017aa}
\bibinfo{author}{\bibfnamefont{F.}~\bibnamefont{Liu}} \bibnamefont{and}
  \bibinfo{author}{\bibfnamefont{K.}~\bibnamefont{Wakabayashi}},
  \bibinfo{journal}{Physical Review Letters} \textbf{\bibinfo{volume}{118}},
  \bibinfo{pages}{076803} (\bibinfo{year}{2017}).

\bibitem[{\citenamefont{Schindler et~al.}(2018)\citenamefont{Schindler, Cook,
  Vergniory, Wang, Parkin, Bernevig, and Neupert}}]{Schindler:2018ab}
\bibinfo{author}{\bibfnamefont{F.}~\bibnamefont{Schindler}},
  \bibinfo{author}{\bibfnamefont{A.~M.} \bibnamefont{Cook}},
  \bibinfo{author}{\bibfnamefont{M.~G.} \bibnamefont{Vergniory}},
  \bibinfo{author}{\bibfnamefont{Z.}~\bibnamefont{Wang}},
  \bibinfo{author}{\bibfnamefont{S.~S.~P.} \bibnamefont{Parkin}},
  \bibinfo{author}{\bibfnamefont{B.~A.} \bibnamefont{Bernevig}},
  \bibnamefont{and} \bibinfo{author}{\bibfnamefont{T.}~\bibnamefont{Neupert}},
  \bibinfo{journal}{Science Advances} \textbf{\bibinfo{volume}{4}},
  \bibinfo{pages}{eaat0346} (\bibinfo{year}{2018}).

\bibitem[{\citenamefont{Langbehn et~al.}(2017)\citenamefont{Langbehn, Peng,
  Trifunovic, von Oppen, and Brouwer}}]{Langbehn:2017aa}
\bibinfo{author}{\bibfnamefont{J.}~\bibnamefont{Langbehn}},
  \bibinfo{author}{\bibfnamefont{Y.}~\bibnamefont{Peng}},
  \bibinfo{author}{\bibfnamefont{L.}~\bibnamefont{Trifunovic}},
  \bibinfo{author}{\bibfnamefont{F.}~\bibnamefont{von Oppen}},
  \bibnamefont{and} \bibinfo{author}{\bibfnamefont{P.~W.}
  \bibnamefont{Brouwer}}, \bibinfo{journal}{Physical Review Letters}
  \textbf{\bibinfo{volume}{119}}, \bibinfo{pages}{246401}
  (\bibinfo{year}{2017}).

\bibitem[{\citenamefont{Song et~al.}(2017)\citenamefont{Song, Fang, and
  Fang}}]{Song:2017aa}
\bibinfo{author}{\bibfnamefont{Z.}~\bibnamefont{Song}},
  \bibinfo{author}{\bibfnamefont{Z.}~\bibnamefont{Fang}}, \bibnamefont{and}
  \bibinfo{author}{\bibfnamefont{C.}~\bibnamefont{Fang}},
  \bibinfo{journal}{Physical Review Letters} \textbf{\bibinfo{volume}{119}},
  \bibinfo{pages}{246402} (\bibinfo{year}{2017}).

\bibitem[{\citenamefont{Khalaf}(2018)}]{Khalaf:2018cr}
\bibinfo{author}{\bibfnamefont{E.}~\bibnamefont{Khalaf}},
  \bibinfo{journal}{Physical Review B} \textbf{\bibinfo{volume}{97}},
  \bibinfo{pages}{205136} (\bibinfo{year}{2018}).

\bibitem[{\citenamefont{Fukui and Hatsugai}(2018)}]{Fukui:2018aa}
\bibinfo{author}{\bibfnamefont{T.}~\bibnamefont{Fukui}} \bibnamefont{and}
  \bibinfo{author}{\bibfnamefont{Y.}~\bibnamefont{Hatsugai}},
  \bibinfo{journal}{Physical Review B} \textbf{\bibinfo{volume}{98}},
  \bibinfo{pages}{035147} (\bibinfo{year}{2018}).

\bibitem[{\citenamefont{Trifunovic and Brouwer}(2019)}]{Trifunovic:2019aa}
\bibinfo{author}{\bibfnamefont{L.}~\bibnamefont{Trifunovic}} \bibnamefont{and}
  \bibinfo{author}{\bibfnamefont{P.~W.} \bibnamefont{Brouwer}},
  \bibinfo{journal}{Physical Review X} \textbf{\bibinfo{volume}{9}},
  \bibinfo{pages}{011012} (\bibinfo{year}{2019}).

\bibitem[{\citenamefont{Benalcazar et~al.}(2019)\citenamefont{Benalcazar, Li,
  and Hughes}}]{Benalcazar:2019aa}
\bibinfo{author}{\bibfnamefont{W.~A.} \bibnamefont{Benalcazar}},
  \bibinfo{author}{\bibfnamefont{T.}~\bibnamefont{Li}}, \bibnamefont{and}
  \bibinfo{author}{\bibfnamefont{T.~L.} \bibnamefont{Hughes}},
  \bibinfo{journal}{Physical Review B} \textbf{\bibinfo{volume}{99}},
  \bibinfo{pages}{245151} (\bibinfo{year}{2019}).

\bibitem[{\citenamefont{Matsugatani and Watanabe}(2018)}]{Matsugatani:2018aa}
\bibinfo{author}{\bibfnamefont{A.}~\bibnamefont{Matsugatani}} \bibnamefont{and}
  \bibinfo{author}{\bibfnamefont{H.}~\bibnamefont{Watanabe}},
  \bibinfo{journal}{Physical Review B} \textbf{\bibinfo{volume}{98}},
  \bibinfo{pages}{205129} (\bibinfo{year}{2018}).

\bibitem[{\citenamefont{Ezawa}(2018{\natexlab{a}})}]{Ezawa:2018aa}
\bibinfo{author}{\bibfnamefont{M.}~\bibnamefont{Ezawa}},
  \bibinfo{journal}{Physical Review Letters} \textbf{\bibinfo{volume}{120}},
  \bibinfo{pages}{026801} (\bibinfo{year}{2018}{\natexlab{a}}).

\bibitem[{\citenamefont{Ezawa}(2018{\natexlab{b}})}]{Ezawa:2018ab}
\bibinfo{author}{\bibfnamefont{M.}~\bibnamefont{Ezawa}},
  \bibinfo{journal}{Physical Review B} \textbf{\bibinfo{volume}{98}},
  \bibinfo{pages}{045125} (\bibinfo{year}{2018}{\natexlab{b}}).

\bibitem[{\citenamefont{C{\u a}lug{\u a}ru et~al.}(2019)\citenamefont{C{\u
  a}lug{\u a}ru, Juri{\v c}i{\'c}, and Roy}}]{Calugaru:2019aa}
\bibinfo{author}{\bibfnamefont{D.}~\bibnamefont{C{\u a}lug{\u a}ru}},
  \bibinfo{author}{\bibfnamefont{V.}~\bibnamefont{Juri{\v c}i{\'c}}},
  \bibnamefont{and} \bibinfo{author}{\bibfnamefont{B.}~\bibnamefont{Roy}},
  \bibinfo{journal}{Physical Review B} \textbf{\bibinfo{volume}{99}},
  \bibinfo{pages}{041301} (\bibinfo{year}{2019}).

\bibitem[{\citenamefont{Ezawa}(2018{\natexlab{c}})}]{Ezawa:2018ac}
\bibinfo{author}{\bibfnamefont{M.}~\bibnamefont{Ezawa}},
  \bibinfo{journal}{Physical Review B} \textbf{\bibinfo{volume}{98}},
  \bibinfo{pages}{201402} (\bibinfo{year}{2018}{\natexlab{c}}).

\bibitem[{\citenamefont{Hashimoto et~al.}(2017)\citenamefont{Hashimoto, Wu, and
  Kimura}}]{Hashimoto:2017aa}
\bibinfo{author}{\bibfnamefont{K.}~\bibnamefont{Hashimoto}},
  \bibinfo{author}{\bibfnamefont{X.}~\bibnamefont{Wu}}, \bibnamefont{and}
  \bibinfo{author}{\bibfnamefont{T.}~\bibnamefont{Kimura}},
  \bibinfo{journal}{Physical Review B} \textbf{\bibinfo{volume}{95}},
  \bibinfo{pages}{165443} (\bibinfo{year}{2017}).

\bibitem[{\citenamefont{Wang et~al.}(2018{\natexlab{a}})\citenamefont{Wang,
  Lin, and Hughes}}]{Wang:2018aa}
\bibinfo{author}{\bibfnamefont{Y.}~\bibnamefont{Wang}},
  \bibinfo{author}{\bibfnamefont{M.}~\bibnamefont{Lin}}, \bibnamefont{and}
  \bibinfo{author}{\bibfnamefont{T.~L.} \bibnamefont{Hughes}},
  \bibinfo{journal}{Physical Review B} \textbf{\bibinfo{volume}{98}},
  \bibinfo{pages}{165144} (\bibinfo{year}{2018}{\natexlab{a}}).

\bibitem[{\citenamefont{Hsu et~al.}(2018)\citenamefont{Hsu, Stano, Klinovaja,
  and Loss}}]{Hsu:2018aa}
\bibinfo{author}{\bibfnamefont{C.-H.} \bibnamefont{Hsu}},
  \bibinfo{author}{\bibfnamefont{P.}~\bibnamefont{Stano}},
  \bibinfo{author}{\bibfnamefont{J.}~\bibnamefont{Klinovaja}},
  \bibnamefont{and} \bibinfo{author}{\bibfnamefont{D.}~\bibnamefont{Loss}},
  \bibinfo{journal}{Physical Review Letters} \textbf{\bibinfo{volume}{121}},
  \bibinfo{pages}{196801} (\bibinfo{year}{2018}).

\bibitem[{\citenamefont{Ghorashi et~al.}(2019)\citenamefont{Ghorashi, Hu,
  Hughes, and Rossi}}]{Ghorashi:2019aa}
\bibinfo{author}{\bibfnamefont{S.~A.~A.} \bibnamefont{Ghorashi}},
  \bibinfo{author}{\bibfnamefont{X.}~\bibnamefont{Hu}},
  \bibinfo{author}{\bibfnamefont{T.~L.} \bibnamefont{Hughes}},
  \bibnamefont{and} \bibinfo{author}{\bibfnamefont{E.}~\bibnamefont{Rossi}},
  \bibinfo{journal}{Physical Review B} \textbf{\bibinfo{volume}{100}},
  \bibinfo{pages}{020509} (\bibinfo{year}{2019}).

\bibitem[{\citenamefont{You et~al.}(2018)\citenamefont{You, Devakul, Burnell,
  and Neupert}}]{You:2018aa}
\bibinfo{author}{\bibfnamefont{Y.}~\bibnamefont{You}},
  \bibinfo{author}{\bibfnamefont{T.}~\bibnamefont{Devakul}},
  \bibinfo{author}{\bibfnamefont{F.~J.} \bibnamefont{Burnell}},
  \bibnamefont{and} \bibinfo{author}{\bibfnamefont{T.}~\bibnamefont{Neupert}},
  \bibinfo{journal}{Physical Review B} \textbf{\bibinfo{volume}{98}},
  \bibinfo{pages}{235102} (\bibinfo{year}{2018}).

\bibitem[{\citenamefont{Bomantara et~al.}(2019)\citenamefont{Bomantara, Zhou,
  Pan, and Gong}}]{Bomantara:2019aa}
\bibinfo{author}{\bibfnamefont{R.~W.} \bibnamefont{Bomantara}},
  \bibinfo{author}{\bibfnamefont{L.}~\bibnamefont{Zhou}},
  \bibinfo{author}{\bibfnamefont{J.}~\bibnamefont{Pan}}, \bibnamefont{and}
  \bibinfo{author}{\bibfnamefont{J.}~\bibnamefont{Gong}},
  \bibinfo{journal}{Physical Review B} \textbf{\bibinfo{volume}{99}},
  \bibinfo{pages}{045441} (\bibinfo{year}{2019}).

\bibitem[{\citenamefont{Rodriguez-Vega
  et~al.}(2018)\citenamefont{Rodriguez-Vega, Kumar, and Seradjeh}}]{1811.04808}
\bibinfo{author}{\bibfnamefont{M.}~\bibnamefont{Rodriguez-Vega}},
  \bibinfo{author}{\bibfnamefont{A.}~\bibnamefont{Kumar}}, \bibnamefont{and}
  \bibinfo{author}{\bibfnamefont{B.}~\bibnamefont{Seradjeh}}
  (\bibinfo{year}{2018}), \eprint{arXiv:1811.04808}.

\bibitem[{\citenamefont{Lieb}(1989)}]{Lieb:1989aa}
\bibinfo{author}{\bibfnamefont{E.~H.} \bibnamefont{Lieb}},
  \bibinfo{journal}{Physical Review Letters} \textbf{\bibinfo{volume}{62}},
  \bibinfo{pages}{1201} (\bibinfo{year}{1989}).

\bibitem[{\citenamefont{Mielke}(1991)}]{Mielke:1991aa}
\bibinfo{author}{\bibfnamefont{A.}~\bibnamefont{Mielke}},
  \textbf{\bibinfo{volume}{24}}, \bibinfo{pages}{L73} (\bibinfo{year}{1991}).

\bibitem[{\citenamefont{Tasaki}(1992)}]{Tasaki:1992aa}
\bibinfo{author}{\bibfnamefont{H.}~\bibnamefont{Tasaki}},
  \bibinfo{journal}{Physical Review Letters} \textbf{\bibinfo{volume}{69}},
  \bibinfo{pages}{1608} (\bibinfo{year}{1992}).

\bibitem[{\citenamefont{Misumi and Aoki}(2017)}]{Misumi:2017aa}
\bibinfo{author}{\bibfnamefont{T.}~\bibnamefont{Misumi}} \bibnamefont{and}
  \bibinfo{author}{\bibfnamefont{H.}~\bibnamefont{Aoki}},
  \bibinfo{journal}{Physical Review B} \textbf{\bibinfo{volume}{96}},
  \bibinfo{pages}{155137} (\bibinfo{year}{2017}).

\bibitem[{\citenamefont{Rhim and Yang}(2019)}]{Rhim:2019aa}
\bibinfo{author}{\bibfnamefont{J.-W.} \bibnamefont{Rhim}} \bibnamefont{and}
  \bibinfo{author}{\bibfnamefont{B.-J.} \bibnamefont{Yang}},
  \bibinfo{journal}{Physical Review B} \textbf{\bibinfo{volume}{99}},
  \bibinfo{pages}{045107} (\bibinfo{year}{2019}).

\bibitem[{\citenamefont{Sun et~al.}(2011)\citenamefont{Sun, Gu, Katsura, and
  Das~Sarma}}]{Sun:2011aa}
\bibinfo{author}{\bibfnamefont{K.}~\bibnamefont{Sun}},
  \bibinfo{author}{\bibfnamefont{Z.}~\bibnamefont{Gu}},
  \bibinfo{author}{\bibfnamefont{H.}~\bibnamefont{Katsura}}, \bibnamefont{and}
  \bibinfo{author}{\bibfnamefont{S.}~\bibnamefont{Das~Sarma}},
  \bibinfo{journal}{Physical Review Letters} \textbf{\bibinfo{volume}{106}},
  \bibinfo{pages}{236803} (\bibinfo{year}{2011}).

\bibitem[{\citenamefont{Tang et~al.}(2011)\citenamefont{Tang, Mei, and
  Wen}}]{Tang:2011aa}
\bibinfo{author}{\bibfnamefont{E.}~\bibnamefont{Tang}},
  \bibinfo{author}{\bibfnamefont{J.-W.} \bibnamefont{Mei}}, \bibnamefont{and}
  \bibinfo{author}{\bibfnamefont{X.-G.} \bibnamefont{Wen}},
  \bibinfo{journal}{Physical Review Letters} \textbf{\bibinfo{volume}{106}},
  \bibinfo{pages}{236802} (\bibinfo{year}{2011}).

\bibitem[{\citenamefont{Neupert et~al.}(2011)\citenamefont{Neupert, Santos,
  Chamon, and Mudry}}]{Neupert:2011aa}
\bibinfo{author}{\bibfnamefont{T.}~\bibnamefont{Neupert}},
  \bibinfo{author}{\bibfnamefont{L.}~\bibnamefont{Santos}},
  \bibinfo{author}{\bibfnamefont{C.}~\bibnamefont{Chamon}}, \bibnamefont{and}
  \bibinfo{author}{\bibfnamefont{C.}~\bibnamefont{Mudry}},
  \bibinfo{journal}{Physical Review Letters} \textbf{\bibinfo{volume}{106}},
  \bibinfo{pages}{236804} (\bibinfo{year}{2011}).

\bibitem[{\citenamefont{Yang et~al.}(2012)\citenamefont{Yang, Gu, Sun, and
  Das~Sarma}}]{Yang:2012aa}
\bibinfo{author}{\bibfnamefont{S.}~\bibnamefont{Yang}},
  \bibinfo{author}{\bibfnamefont{Z.-C.} \bibnamefont{Gu}},
  \bibinfo{author}{\bibfnamefont{K.}~\bibnamefont{Sun}}, \bibnamefont{and}
  \bibinfo{author}{\bibfnamefont{S.}~\bibnamefont{Das~Sarma}},
  \bibinfo{journal}{Physical Review B} \textbf{\bibinfo{volume}{86}},
  \bibinfo{pages}{241112} (\bibinfo{year}{2012}).

\bibitem[{\citenamefont{Kobayashi et~al.}(2016)\citenamefont{Kobayashi,
  Okumura, Yamada, Machida, and Aoki}}]{Kobayashi:2016aa}
\bibinfo{author}{\bibfnamefont{K.}~\bibnamefont{Kobayashi}},
  \bibinfo{author}{\bibfnamefont{M.}~\bibnamefont{Okumura}},
  \bibinfo{author}{\bibfnamefont{S.}~\bibnamefont{Yamada}},
  \bibinfo{author}{\bibfnamefont{M.}~\bibnamefont{Machida}}, \bibnamefont{and}
  \bibinfo{author}{\bibfnamefont{H.}~\bibnamefont{Aoki}},
  \bibinfo{journal}{Physical Review B} \textbf{\bibinfo{volume}{94}},
  \bibinfo{pages}{214501} (\bibinfo{year}{2016}).

\bibitem[{\citenamefont{Tovmasyan et~al.}(2016)\citenamefont{Tovmasyan, Peotta,
  T{\"o}rm{\"a}, and Huber}}]{Tovmasyan:2016aa}
\bibinfo{author}{\bibfnamefont{M.}~\bibnamefont{Tovmasyan}},
  \bibinfo{author}{\bibfnamefont{S.}~\bibnamefont{Peotta}},
  \bibinfo{author}{\bibfnamefont{P.}~\bibnamefont{T{\"o}rm{\"a}}},
  \bibnamefont{and} \bibinfo{author}{\bibfnamefont{S.~D.} \bibnamefont{Huber}},
  \bibinfo{journal}{Physical Review B} \textbf{\bibinfo{volume}{94}},
  \bibinfo{pages}{245149} (\bibinfo{year}{2016}).

\bibitem[{\citenamefont{Cao et~al.}(2018)\citenamefont{Cao, Fatemi, Fang,
  Watanabe, Taniguchi, Kaxiras, and Jarillo-Herrero}}]{Cao:2018aa}
\bibinfo{author}{\bibfnamefont{Y.}~\bibnamefont{Cao}},
  \bibinfo{author}{\bibfnamefont{V.}~\bibnamefont{Fatemi}},
  \bibinfo{author}{\bibfnamefont{S.}~\bibnamefont{Fang}},
  \bibinfo{author}{\bibfnamefont{K.}~\bibnamefont{Watanabe}},
  \bibinfo{author}{\bibfnamefont{T.}~\bibnamefont{Taniguchi}},
  \bibinfo{author}{\bibfnamefont{E.}~\bibnamefont{Kaxiras}}, \bibnamefont{and}
  \bibinfo{author}{\bibfnamefont{P.}~\bibnamefont{Jarillo-Herrero}},
  \bibinfo{journal}{Nature} \textbf{\bibinfo{volume}{556}}, \bibinfo{pages}{43
  EP } (\bibinfo{year}{2018}).

\bibitem[{\citenamefont{Fatemi et~al.}(2018)\citenamefont{Fatemi, Wu, Cao,
  Bretheau, Gibson, Watanabe, Taniguchi, Cava, and
  Jarillo-Herrero}}]{Fatemi:2018aa}
\bibinfo{author}{\bibfnamefont{V.}~\bibnamefont{Fatemi}},
  \bibinfo{author}{\bibfnamefont{S.}~\bibnamefont{Wu}},
  \bibinfo{author}{\bibfnamefont{Y.}~\bibnamefont{Cao}},
  \bibinfo{author}{\bibfnamefont{L.}~\bibnamefont{Bretheau}},
  \bibinfo{author}{\bibfnamefont{Q.~D.} \bibnamefont{Gibson}},
  \bibinfo{author}{\bibfnamefont{K.}~\bibnamefont{Watanabe}},
  \bibinfo{author}{\bibfnamefont{T.}~\bibnamefont{Taniguchi}},
  \bibinfo{author}{\bibfnamefont{R.~J.} \bibnamefont{Cava}}, \bibnamefont{and}
  \bibinfo{author}{\bibfnamefont{P.}~\bibnamefont{Jarillo-Herrero}},
  \bibinfo{journal}{Science} \textbf{\bibinfo{volume}{362}},
  \bibinfo{pages}{926} (\bibinfo{year}{2018}).

\bibitem[{\citenamefont{Zou et~al.}(2018)\citenamefont{Zou, Po, Vishwanath, and
  Senthil}}]{Zou:2018aa}
\bibinfo{author}{\bibfnamefont{L.}~\bibnamefont{Zou}},
  \bibinfo{author}{\bibfnamefont{H.~C.} \bibnamefont{Po}},
  \bibinfo{author}{\bibfnamefont{A.}~\bibnamefont{Vishwanath}},
  \bibnamefont{and} \bibinfo{author}{\bibfnamefont{T.}~\bibnamefont{Senthil}},
  \bibinfo{journal}{Physical Review B} \textbf{\bibinfo{volume}{98}},
  \bibinfo{pages}{085435} (\bibinfo{year}{2018}).

\bibitem[{\citenamefont{Rademaker and Mellado}(2018)}]{Rademaker:2018aa}
\bibinfo{author}{\bibfnamefont{L.}~\bibnamefont{Rademaker}} \bibnamefont{and}
  \bibinfo{author}{\bibfnamefont{P.}~\bibnamefont{Mellado}},
  \bibinfo{journal}{Physical Review B} \textbf{\bibinfo{volume}{98}},
  \bibinfo{pages}{235158} (\bibinfo{year}{2018}).

\bibitem[{\citenamefont{Venderbos and Fernandes}(2018)}]{Venderbos:2018aa}
\bibinfo{author}{\bibfnamefont{J.~W.~F.} \bibnamefont{Venderbos}}
  \bibnamefont{and} \bibinfo{author}{\bibfnamefont{R.~M.}
  \bibnamefont{Fernandes}}, \bibinfo{journal}{Physical Review B}
  \textbf{\bibinfo{volume}{98}}, \bibinfo{pages}{245103}
  (\bibinfo{year}{2018}).

\bibitem[{\citenamefont{Ramires and Lado}(2018)}]{Ramires:2018aa}
\bibinfo{author}{\bibfnamefont{A.}~\bibnamefont{Ramires}} \bibnamefont{and}
  \bibinfo{author}{\bibfnamefont{J.~L.} \bibnamefont{Lado}},
  \bibinfo{journal}{Physical Review Letters} \textbf{\bibinfo{volume}{121}},
  \bibinfo{pages}{146801} (\bibinfo{year}{2018}).

\bibitem[{\citenamefont{Koshino et~al.}(2018)\citenamefont{Koshino, Yuan,
  Koretsune, Ochi, Kuroki, and Fu}}]{Koshino:2018aa}
\bibinfo{author}{\bibfnamefont{M.}~\bibnamefont{Koshino}},
  \bibinfo{author}{\bibfnamefont{N.~F.~Q.} \bibnamefont{Yuan}},
  \bibinfo{author}{\bibfnamefont{T.}~\bibnamefont{Koretsune}},
  \bibinfo{author}{\bibfnamefont{M.}~\bibnamefont{Ochi}},
  \bibinfo{author}{\bibfnamefont{K.}~\bibnamefont{Kuroki}}, \bibnamefont{and}
  \bibinfo{author}{\bibfnamefont{L.}~\bibnamefont{Fu}},
  \bibinfo{journal}{Physical Review X} \textbf{\bibinfo{volume}{8}},
  \bibinfo{pages}{031087} (\bibinfo{year}{2018}).

\bibitem[{\citenamefont{Peltonen et~al.}(2018)\citenamefont{Peltonen,
  Ojaj{\"a}rvi, and Heikkil{\"a}}}]{Peltonen:2018aa}
\bibinfo{author}{\bibfnamefont{T.~J.} \bibnamefont{Peltonen}},
  \bibinfo{author}{\bibfnamefont{R.}~\bibnamefont{Ojaj{\"a}rvi}},
  \bibnamefont{and} \bibinfo{author}{\bibfnamefont{T.~T.}
  \bibnamefont{Heikkil{\"a}}}, \bibinfo{journal}{Physical Review B}
  \textbf{\bibinfo{volume}{98}}, \bibinfo{pages}{220504}
  (\bibinfo{year}{2018}).

\bibitem[{\citenamefont{Po et~al.}(2018)\citenamefont{Po, Zou, Vishwanath, and
  Senthil}}]{Po:2018aa}
\bibinfo{author}{\bibfnamefont{H.~C.} \bibnamefont{Po}},
  \bibinfo{author}{\bibfnamefont{L.}~\bibnamefont{Zou}},
  \bibinfo{author}{\bibfnamefont{A.}~\bibnamefont{Vishwanath}},
  \bibnamefont{and} \bibinfo{author}{\bibfnamefont{T.}~\bibnamefont{Senthil}},
  \bibinfo{journal}{Physical Review X} \textbf{\bibinfo{volume}{8}},
  \bibinfo{pages}{031089} (\bibinfo{year}{2018}).

\bibitem[{\citenamefont{Guo et~al.}(2018)\citenamefont{Guo, Zhu, Feng, and
  Scalettar}}]{Guo:2018aa}
\bibinfo{author}{\bibfnamefont{H.}~\bibnamefont{Guo}},
  \bibinfo{author}{\bibfnamefont{X.}~\bibnamefont{Zhu}},
  \bibinfo{author}{\bibfnamefont{S.}~\bibnamefont{Feng}}, \bibnamefont{and}
  \bibinfo{author}{\bibfnamefont{R.~T.} \bibnamefont{Scalettar}},
  \bibinfo{journal}{Physical Review B} \textbf{\bibinfo{volume}{97}},
  \bibinfo{pages}{235453} (\bibinfo{year}{2018}).

\bibitem[{\citenamefont{Ochi et~al.}(2018)\citenamefont{Ochi, Koshino, and
  Kuroki}}]{Ochi:2018aa}
\bibinfo{author}{\bibfnamefont{M.}~\bibnamefont{Ochi}},
  \bibinfo{author}{\bibfnamefont{M.}~\bibnamefont{Koshino}}, \bibnamefont{and}
  \bibinfo{author}{\bibfnamefont{K.}~\bibnamefont{Kuroki}},
  \bibinfo{journal}{Physical Review B} \textbf{\bibinfo{volume}{98}},
  \bibinfo{pages}{081102} (\bibinfo{year}{2018}).

\bibitem[{\citenamefont{Lin et~al.}(2018)\citenamefont{Lin, Liu, and
  Tom{\'a}nek}}]{Lin:2018aa}
\bibinfo{author}{\bibfnamefont{X.}~\bibnamefont{Lin}},
  \bibinfo{author}{\bibfnamefont{D.}~\bibnamefont{Liu}}, \bibnamefont{and}
  \bibinfo{author}{\bibfnamefont{D.}~\bibnamefont{Tom{\'a}nek}},
  \bibinfo{journal}{Physical Review B} \textbf{\bibinfo{volume}{98}},
  \bibinfo{pages}{195432} (\bibinfo{year}{2018}).

\bibitem[{\citenamefont{Kennes et~al.}(2018)\citenamefont{Kennes, Lischner, and
  Karrasch}}]{Kennes:2018aa}
\bibinfo{author}{\bibfnamefont{D.~M.} \bibnamefont{Kennes}},
  \bibinfo{author}{\bibfnamefont{J.}~\bibnamefont{Lischner}}, \bibnamefont{and}
  \bibinfo{author}{\bibfnamefont{C.}~\bibnamefont{Karrasch}},
  \bibinfo{journal}{Physical Review B} \textbf{\bibinfo{volume}{98}},
  \bibinfo{pages}{241407} (\bibinfo{year}{2018}).

\bibitem[{\citenamefont{Choi and Choi}(2018)}]{Choi:2018aa}
\bibinfo{author}{\bibfnamefont{Y.~W.} \bibnamefont{Choi}} \bibnamefont{and}
  \bibinfo{author}{\bibfnamefont{H.~J.} \bibnamefont{Choi}},
  \bibinfo{journal}{Physical Review B} \textbf{\bibinfo{volume}{98}},
  \bibinfo{pages}{241412} (\bibinfo{year}{2018}).

\bibitem[{\citenamefont{Qiao et~al.}(2018)\citenamefont{Qiao, Yin, and
  He}}]{Qiao:2018aa}
\bibinfo{author}{\bibfnamefont{J.-B.} \bibnamefont{Qiao}},
  \bibinfo{author}{\bibfnamefont{L.-J.} \bibnamefont{Yin}}, \bibnamefont{and}
  \bibinfo{author}{\bibfnamefont{L.}~\bibnamefont{He}},
  \bibinfo{journal}{Physical Review B} \textbf{\bibinfo{volume}{98}},
  \bibinfo{pages}{235402} (\bibinfo{year}{2018}).

\bibitem[{\citenamefont{Hejazi et~al.}(2019)\citenamefont{Hejazi, Liu,
  Shapourian, Chen, and Balents}}]{Hejazi:2019aa}
\bibinfo{author}{\bibfnamefont{K.}~\bibnamefont{Hejazi}},
  \bibinfo{author}{\bibfnamefont{C.}~\bibnamefont{Liu}},
  \bibinfo{author}{\bibfnamefont{H.}~\bibnamefont{Shapourian}},
  \bibinfo{author}{\bibfnamefont{X.}~\bibnamefont{Chen}}, \bibnamefont{and}
  \bibinfo{author}{\bibfnamefont{L.}~\bibnamefont{Balents}},
  \bibinfo{journal}{Physical Review B} \textbf{\bibinfo{volume}{99}},
  \bibinfo{pages}{035111} (\bibinfo{year}{2019}).

\bibitem[{\citenamefont{Tarnopolsky et~al.}(2019)\citenamefont{Tarnopolsky,
  Kruchkov, and Vishwanath}}]{Tarnopolsky:2019aa}
\bibinfo{author}{\bibfnamefont{G.}~\bibnamefont{Tarnopolsky}},
  \bibinfo{author}{\bibfnamefont{A.~J.} \bibnamefont{Kruchkov}},
  \bibnamefont{and}
  \bibinfo{author}{\bibfnamefont{A.}~\bibnamefont{Vishwanath}},
  \bibinfo{journal}{Physical Review Letters} \textbf{\bibinfo{volume}{122}},
  \bibinfo{pages}{106405} (\bibinfo{year}{2019}).

\bibitem[{\citenamefont{Yu et~al.}(2015)\citenamefont{Yu, Weng, Fang, Dai, and
  Hu}}]{Yu:2015aa}
\bibinfo{author}{\bibfnamefont{R.}~\bibnamefont{Yu}},
  \bibinfo{author}{\bibfnamefont{H.}~\bibnamefont{Weng}},
  \bibinfo{author}{\bibfnamefont{Z.}~\bibnamefont{Fang}},
  \bibinfo{author}{\bibfnamefont{X.}~\bibnamefont{Dai}}, \bibnamefont{and}
  \bibinfo{author}{\bibfnamefont{X.}~\bibnamefont{Hu}},
  \bibinfo{journal}{Physical Review Letters} \textbf{\bibinfo{volume}{115}},
  \bibinfo{pages}{036807} (\bibinfo{year}{2015}).

\bibitem[{\citenamefont{Kim et~al.}(2015)\citenamefont{Kim, Wieder, Kane, and
  Rappe}}]{Kim:2015aa}
\bibinfo{author}{\bibfnamefont{Y.}~\bibnamefont{Kim}},
  \bibinfo{author}{\bibfnamefont{B.~J.} \bibnamefont{Wieder}},
  \bibinfo{author}{\bibfnamefont{C.~L.} \bibnamefont{Kane}}, \bibnamefont{and}
  \bibinfo{author}{\bibfnamefont{A.~M.} \bibnamefont{Rappe}},
  \bibinfo{journal}{Physical Review Letters} \textbf{\bibinfo{volume}{115}},
  \bibinfo{pages}{036806} (\bibinfo{year}{2015}).

\bibitem[{\citenamefont{Huang et~al.}(2016)\citenamefont{Huang, Liu,
  Vanderbilt, and Duan}}]{Huang:2016aa}
\bibinfo{author}{\bibfnamefont{H.}~\bibnamefont{Huang}},
  \bibinfo{author}{\bibfnamefont{J.}~\bibnamefont{Liu}},
  \bibinfo{author}{\bibfnamefont{D.}~\bibnamefont{Vanderbilt}},
  \bibnamefont{and} \bibinfo{author}{\bibfnamefont{W.}~\bibnamefont{Duan}},
  \bibinfo{journal}{Physical Review B} \textbf{\bibinfo{volume}{93}},
  \bibinfo{pages}{201114} (\bibinfo{year}{2016}).

\bibitem[{\citenamefont{Zhang et~al.}(2016)\citenamefont{Zhang, Zhao, Liu, Xue,
  Zhu, and Wang}}]{Zhang:2016ab}
\bibinfo{author}{\bibfnamefont{D.-W.} \bibnamefont{Zhang}},
  \bibinfo{author}{\bibfnamefont{Y.~X.} \bibnamefont{Zhao}},
  \bibinfo{author}{\bibfnamefont{R.-B.} \bibnamefont{Liu}},
  \bibinfo{author}{\bibfnamefont{Z.-Y.} \bibnamefont{Xue}},
  \bibinfo{author}{\bibfnamefont{S.-L.} \bibnamefont{Zhu}}, \bibnamefont{and}
  \bibinfo{author}{\bibfnamefont{Z.~D.} \bibnamefont{Wang}},
  \bibinfo{journal}{Physical Review A} \textbf{\bibinfo{volume}{93}},
  \bibinfo{pages}{043617} (\bibinfo{year}{2016}).

\bibitem[{\citenamefont{Chan et~al.}(2016)\citenamefont{Chan, Chiu, Chou, and
  Schnyder}}]{Chan:2016aa}
\bibinfo{author}{\bibfnamefont{Y.~H.} \bibnamefont{Chan}},
  \bibinfo{author}{\bibfnamefont{C.-K.} \bibnamefont{Chiu}},
  \bibinfo{author}{\bibfnamefont{M.~Y.} \bibnamefont{Chou}}, \bibnamefont{and}
  \bibinfo{author}{\bibfnamefont{A.~P.} \bibnamefont{Schnyder}},
  \bibinfo{journal}{Physical Review B} \textbf{\bibinfo{volume}{93}},
  \bibinfo{pages}{205132} (\bibinfo{year}{2016}).

\bibitem[{\citenamefont{Pal and Saha}(2018)}]{Pal:2018ab}
\bibinfo{author}{\bibfnamefont{B.}~\bibnamefont{Pal}} \bibnamefont{and}
  \bibinfo{author}{\bibfnamefont{K.}~\bibnamefont{Saha}},
  \bibinfo{journal}{Physical Review B} \textbf{\bibinfo{volume}{97}},
  \bibinfo{pages}{195101} (\bibinfo{year}{2018}).

\bibitem[{\citenamefont{Pal}(2018)}]{Pal:2018aa}
\bibinfo{author}{\bibfnamefont{B.}~\bibnamefont{Pal}},
  \bibinfo{journal}{Physical Review B} \textbf{\bibinfo{volume}{98}},
  \bibinfo{pages}{245116} (\bibinfo{year}{2018}).

\bibitem[{\citenamefont{Li et~al.}(2018{\natexlab{a}})\citenamefont{Li, Liu,
  Fu, Yu, Yang, and Yao}}]{Li:2018aa}
\bibinfo{author}{\bibfnamefont{S.}~\bibnamefont{Li}},
  \bibinfo{author}{\bibfnamefont{Y.}~\bibnamefont{Liu}},
  \bibinfo{author}{\bibfnamefont{B.}~\bibnamefont{Fu}},
  \bibinfo{author}{\bibfnamefont{Z.-M.} \bibnamefont{Yu}},
  \bibinfo{author}{\bibfnamefont{S.~A.} \bibnamefont{Yang}}, \bibnamefont{and}
  \bibinfo{author}{\bibfnamefont{Y.}~\bibnamefont{Yao}},
  \bibinfo{journal}{Physical Review B} \textbf{\bibinfo{volume}{97}},
  \bibinfo{pages}{245148} (\bibinfo{year}{2018}{\natexlab{a}}).

\bibitem[{\citenamefont{Ahn et~al.}(2018{\natexlab{a}})\citenamefont{Ahn, Kim,
  Kim, and Yang}}]{Ahn:2018aa}
\bibinfo{author}{\bibfnamefont{J.}~\bibnamefont{Ahn}},
  \bibinfo{author}{\bibfnamefont{D.}~\bibnamefont{Kim}},
  \bibinfo{author}{\bibfnamefont{Y.}~\bibnamefont{Kim}}, \bibnamefont{and}
  \bibinfo{author}{\bibfnamefont{B.-J.} \bibnamefont{Yang}},
  \bibinfo{journal}{Physical Review Letters} \textbf{\bibinfo{volume}{121}},
  \bibinfo{pages}{106403} (\bibinfo{year}{2018}{\natexlab{a}}).

\bibitem[{\citenamefont{Wang et~al.}(2018{\natexlab{b}})\citenamefont{Wang,
  Wieder, Li, Yan, and Bernevig}}]{1806.11116}
\bibinfo{author}{\bibfnamefont{Z.}~\bibnamefont{Wang}},
  \bibinfo{author}{\bibfnamefont{B.~J.} \bibnamefont{Wieder}},
  \bibinfo{author}{\bibfnamefont{J.}~\bibnamefont{Li}},
  \bibinfo{author}{\bibfnamefont{B.}~\bibnamefont{Yan}}, \bibnamefont{and}
  \bibinfo{author}{\bibfnamefont{B.~A.} \bibnamefont{Bernevig}}
  (\bibinfo{year}{2018}{\natexlab{b}}), \eprint{arXiv:1806.11116}.

\bibitem[{\citenamefont{Ahn et~al.}(2018{\natexlab{b}})\citenamefont{Ahn, Park,
  and Yang}}]{1808.05375}
\bibinfo{author}{\bibfnamefont{J.}~\bibnamefont{Ahn}},
  \bibinfo{author}{\bibfnamefont{S.}~\bibnamefont{Park}}, \bibnamefont{and}
  \bibinfo{author}{\bibfnamefont{B.-J.} \bibnamefont{Yang}}
  (\bibinfo{year}{2018}{\natexlab{b}}), \eprint{arXiv:1808.05375}.

\bibitem[{\citenamefont{Wieder and Bernevig}(2018)}]{1810.02373}
\bibinfo{author}{\bibfnamefont{B.~J.} \bibnamefont{Wieder}} \bibnamefont{and}
  \bibinfo{author}{\bibfnamefont{B.~A.} \bibnamefont{Bernevig}}
  (\bibinfo{year}{2018}), \eprint{arXiv:1810.02373}.

\bibitem[{\citenamefont{Chiu et~al.}(2018)\citenamefont{Chiu, Chan, and
  Schnyder}}]{1810.04094}
\bibinfo{author}{\bibfnamefont{C.~K.} \bibnamefont{Chiu}},
  \bibinfo{author}{\bibfnamefont{Y.~H.} \bibnamefont{Chan}}, \bibnamefont{and}
  \bibinfo{author}{\bibfnamefont{A.~P.} \bibnamefont{Schnyder}}
  (\bibinfo{year}{2018}), \eprint{arXiv:1810.04094}.

\bibitem[{\citenamefont{Ahn and Yang}(2019)}]{Ahn:2019aa}
\bibinfo{author}{\bibfnamefont{J.}~\bibnamefont{Ahn}} \bibnamefont{and}
  \bibinfo{author}{\bibfnamefont{B.-J.} \bibnamefont{Yang}},
  \bibinfo{journal}{Physical Review B} \textbf{\bibinfo{volume}{99}},
  \bibinfo{pages}{235125} (\bibinfo{year}{2019}).

\bibitem[{\citenamefont{Ghatak and Das}(2019)}]{Ghatak_2019}
\bibinfo{author}{\bibfnamefont{A.}~\bibnamefont{Ghatak}} \bibnamefont{and}
  \bibinfo{author}{\bibfnamefont{T.}~\bibnamefont{Das}},
  \bibinfo{journal}{Journal of Physics: Condensed Matter}
  \textbf{\bibinfo{volume}{31}}, \bibinfo{pages}{263001}
  (\bibinfo{year}{2019}).

\bibitem[{\citenamefont{Su et~al.}(1979)\citenamefont{Su, Schrieffer, and
  Heeger}}]{Su:1979aa}
\bibinfo{author}{\bibfnamefont{W.~P.} \bibnamefont{Su}},
  \bibinfo{author}{\bibfnamefont{J.~R.} \bibnamefont{Schrieffer}},
  \bibnamefont{and} \bibinfo{author}{\bibfnamefont{A.~J.}
  \bibnamefont{Heeger}}, \bibinfo{journal}{Physical Review Letters}
  \textbf{\bibinfo{volume}{42}}, \bibinfo{pages}{1698} (\bibinfo{year}{1979}).

\bibitem[{\citenamefont{Haldane}(1988)}]{Haldane:1988aa}
\bibinfo{author}{\bibfnamefont{F.~D.~M.} \bibnamefont{Haldane}},
  \bibinfo{journal}{Physical Review Letters} \textbf{\bibinfo{volume}{61}},
  \bibinfo{pages}{2015} (\bibinfo{year}{1988}).

\bibitem[{\citenamefont{Fukui et~al.}(2005)\citenamefont{Fukui, Hatsugai, and
  Suzuki}}]{FHS05}
\bibinfo{author}{\bibfnamefont{T.}~\bibnamefont{Fukui}},
  \bibinfo{author}{\bibfnamefont{Y.}~\bibnamefont{Hatsugai}}, \bibnamefont{and}
  \bibinfo{author}{\bibfnamefont{H.}~\bibnamefont{Suzuki}},
  \bibinfo{journal}{Journal of the Physical Society of Japan}
  \textbf{\bibinfo{volume}{74}}, \bibinfo{pages}{1674} (\bibinfo{year}{2005}).

\bibitem[{\citenamefont{Fang et~al.}(2012)\citenamefont{Fang, Gilbert, and
  Bernevig}}]{Fang:2012aa}
\bibinfo{author}{\bibfnamefont{C.}~\bibnamefont{Fang}},
  \bibinfo{author}{\bibfnamefont{M.~J.} \bibnamefont{Gilbert}},
  \bibnamefont{and} \bibinfo{author}{\bibfnamefont{B.~A.}
  \bibnamefont{Bernevig}}, \bibinfo{journal}{Physical Review B}
  \textbf{\bibinfo{volume}{86}}, \bibinfo{pages}{115112}
  (\bibinfo{year}{2012}).

\bibitem[{\citenamefont{Li et~al.}(2018{\natexlab{b}})\citenamefont{Li, Umer,
  and Gong}}]{Li:2018ab}
\bibinfo{author}{\bibfnamefont{L.}~\bibnamefont{Li}},
  \bibinfo{author}{\bibfnamefont{M.}~\bibnamefont{Umer}}, \bibnamefont{and}
  \bibinfo{author}{\bibfnamefont{J.}~\bibnamefont{Gong}},
  \bibinfo{journal}{Physical Review B} \textbf{\bibinfo{volume}{98}},
  \bibinfo{pages}{205422} (\bibinfo{year}{2018}{\natexlab{b}}).

\end{thebibliography}

\end{document}